\documentclass[twocolumn]{aastex63}
\usepackage{bm}
\shorttitle{The pulsar gamma-ray emission}
\shortauthors{Cao \& Yang}
\begin{document}
\title{The pulsar gamma-ray emission from the high-resolution dissipative  magnetospheres}
\author{Gang Cao}
\author{Xiongbang Yang}
\affiliation{Department of  Mathematics, Yunnan University of Finance and Economics, Kunming 650221, Yunnan, P. R. China, gcao@ynufe.edu.cn}
\begin{abstract}

The pulsar light curves and energy spectra are  explored in dissipative pulsar magnetospheres  with the Aristotelian electrodynamics (AE), where  particle acceleration is fully balanced with  radiation reaction. The  AE magnetospheres with non-zero pair multiplicity  are  computed   by a pseudo-spectral method in the co-moving frame.
The dissipative region near the current sheet outside the LC is accurately captured by the high-resolution simulation.
The pulsar light curves and spectra are computed by the test particle trajectory method including the influence of both the consistent accelerating electric field and radiation reaction. Our results  can generally reproduce the double-peak light curves and the GeV cut-off energy spectra in agreement with the Fermi observations for the pair multiplicity $\kappa\gtrsim1$.
\end{abstract}


\section{Introduction}
The Fermi Gamma-Ray Space Telescope launched in 2008 has opened a new era  on the study of pulsar $\gamma$-ray emission. To data, the Fermi LAT have detected more than 270  $\gamma$-ray pulsars\footnote{https://confluence.slac.stanford.edu/display/GLAMCOG/ \\ Public+List+of+LAT-Detected+Gamma-Ray+Pulsars}, 117 of which are listed in the Second Fermi Pulsar Catalog \citep{abd10,abd13}. Fermi $\gamma$-ray pulsar are classified into three groups: young radio-loud, young radio-quiet and millisecond pulsars. The light curves from these pulsars  usually show the widely separated double-peak profiles, and the first peak lags the radio peak by a small fraction of rotation period. The pulsar $\gamma$-ray  spectra can be described by a power law with an exponential cut-off, and  the cut-off energy is in the range of 1-5 GeV. The light curves and energy spectra from the Fermi LAT offer a unique insight to explore the  nature of the radiation mechanisms and the location of particle acceleration  in the magnetosphere. However, it is still uncertain about the origin of pulsar emission. In fact, the pulsed emission and particle acceleration in the magnetosphere are  controlled by the structure of the global pulsar  magnetosphere. This requires us to have a deep and accurate knowledge of the  pulsar magnetosphere. The significant progresses  have been made in the numerical model of the global pulsar magnetospheres within the last decades.

It is well believed that the  pulsar magnetosphere is loaded with plasmas by  pair creation\citep{gol69}. A zeroth order approximation of the plasma-filled magnetosphere is referred to the force-free electrodynamics(FFE). The force-free approximation requires the density number much larger than GJ density, these plasmas quickly short out the accelerating fields so that the force-free condition ${\bf E}\cdot {\bf B}=0$ holds everywhere. The force-free pulsar magnetosphere have been recently achieved with the advent of numerical simulations. The numerical force-free solution is firstly obtained by \citet{con99} for the aligned rotator and then by \citet{spi06} for the oblique rotator. 3D force-free solutions are  further explored by the finite-difference time-domain method \citep{kal09,con10} and the spectral method \citep{pet12,cao16b,pet16}. All these time-dependent force-free simulations confirmed  the existence of the current sheet outside the LC, which is thought as the potential site of the pulsar $\gamma$-ray emission. The pulsar light curves and spectra are also studied by placing the emission region in the current sheet outside the LC \citep{bai10,har15,bra15,bog18,har18}. However, the force-free solution do not allow any particle acceleration and production of the pulsed radiation in the magnetosphere.

More realistic pulsar model should  allow the local dissipation to produce the observed pulsar phenomenons. The dissipative effects have been included by involving a finite conductivity \citep{li12,kal12a,cao16b}, which is called the resistive magnetosphere. The resistive magnetosphere ranges  from the vacuum limit to force-free limit with increasing conductivity, and the resistive solution  produces accelerating electric fields that are self-consistent with the magnetic field structure. The resistive magnetospheres have been used to model the pulsar light curves \citep{kal12b,kal14,kal17,cao19} and energy spectra \citep{yang21} by including the accelerating electric field from the simulation. These studies suggested that the particle acceleration and the $\gamma$-ray radiation is produced near the current sheets outside the LC. Recently, particle-in-cell (PIC) methods  are developed to model the pulsar magnetosphere by self-consistently treating the  feedback between particle motions and the electromagnetic fields \citep{phi14,che14,phi15,bel15,cer15,kal18,bra18}. Moreover, the pulsar light curves are predicted by extracting particle radiation along each trajectory in full PIC simualtion \citep{cer16,phi18,kal18}. However,  the particle energy from the PIC simulation are much small than those in the real pulsar, which is  not enough to produce the observed Fermi $\gamma$-ray emission.\\

A good approximation between the resistive model and PIC model is Aristotelian electrodynamics, which can include the back-reaction of the emitting photons onto particle motions and allow
for some dissipations in the magnetosphere. The AE method is firstly introduced to study the pulsar magnetosphere by \citet{gru12,gru13}.
Recently, a clever method by combining FFE with AE is proposed to construct the structure of pulsar magnetosphere \citep{con16,pet20a,cao20}. \citet{con16}
first presented the 3D structure of the AE magnetosphere for the oblique rotator by the finite-difference method. However, they only studied the  AE solution in the limit of  no pair multiplicity. \citet{pet20a} extended the study of \citet{con16} by introducing non-zero pair multiplicity  and used the spectral method to compute the AE magnetosphere but only for the aligned rotator. Recently,
\citet{cao20} presented the first AE solution with non-zero pair multiplicity  for the oblique rotator by the spectral method. They show that the dissipative region is more restricted to the current sheet outside LC as the pair multiplicity increases. However, a relatively low resolution is generally used in all these simulations, which is not enough to accurately capture the current sheet outside the LC.  Moreover,  the light curves and spectra were not computed from the numerical AE solutions in all these studies. In this paper, we  present the high-resolution
simulation of the AE magnetosphere with non-zero pair multiplicity for the oblique rotator by the spectral method. The pulsar light curves and spectra are then computed  by the test particle trajectory method in dissipative AE magnetosphere.
The paper is organized as follows:  We describe the  AE model in Section 2. We present the pulsar light curves and spectra from the AE magnetosphere in Section 3. A brief discussion and conclusions are presented in Section 4.\\
\begin{figure*}
\center
\begin{tabular}{cccccccccccccc}
\includegraphics[width=5.85 cm,height=5.5 cm]{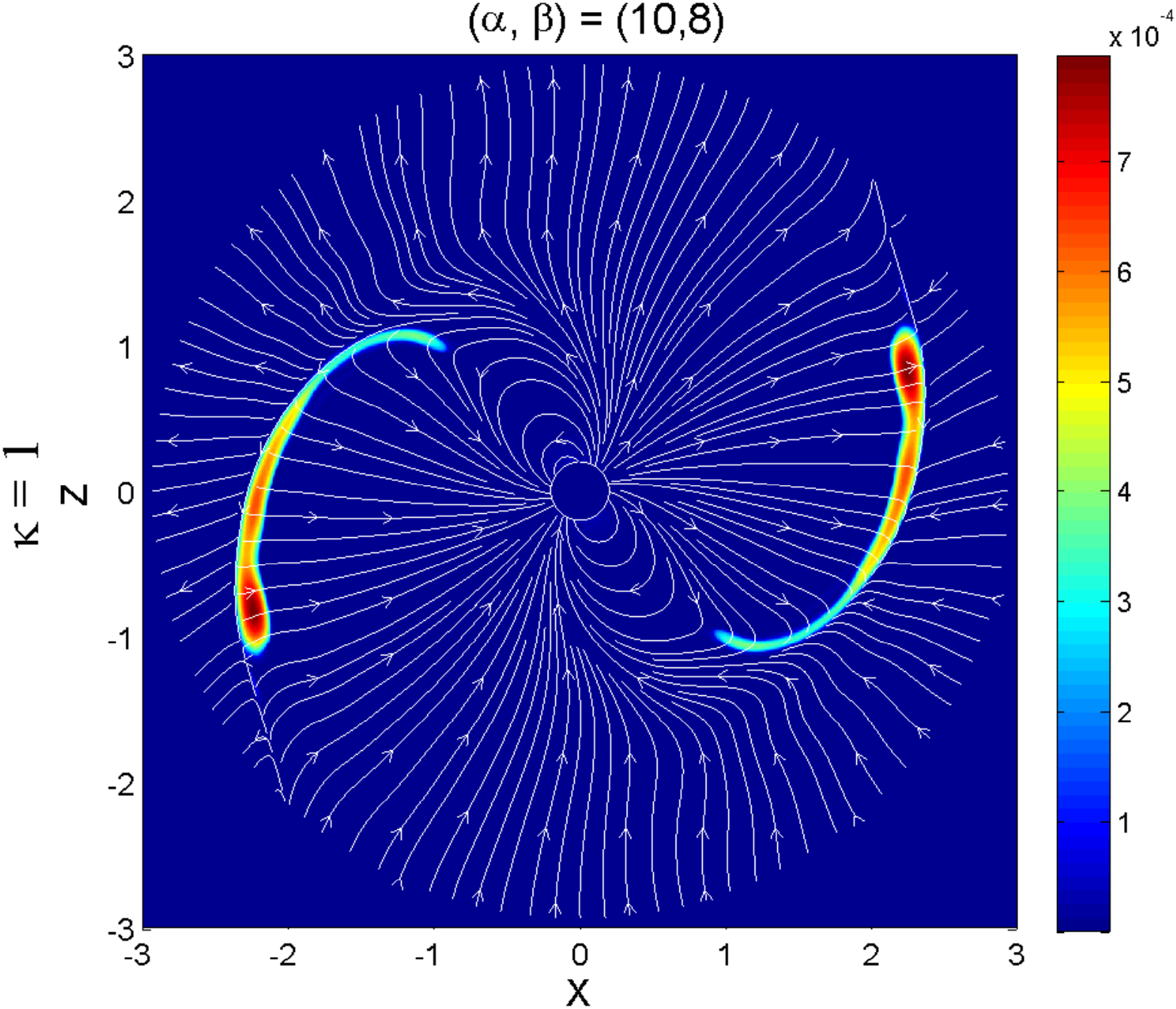} \qquad
\includegraphics[width=5.4 cm,height=5.5 cm]{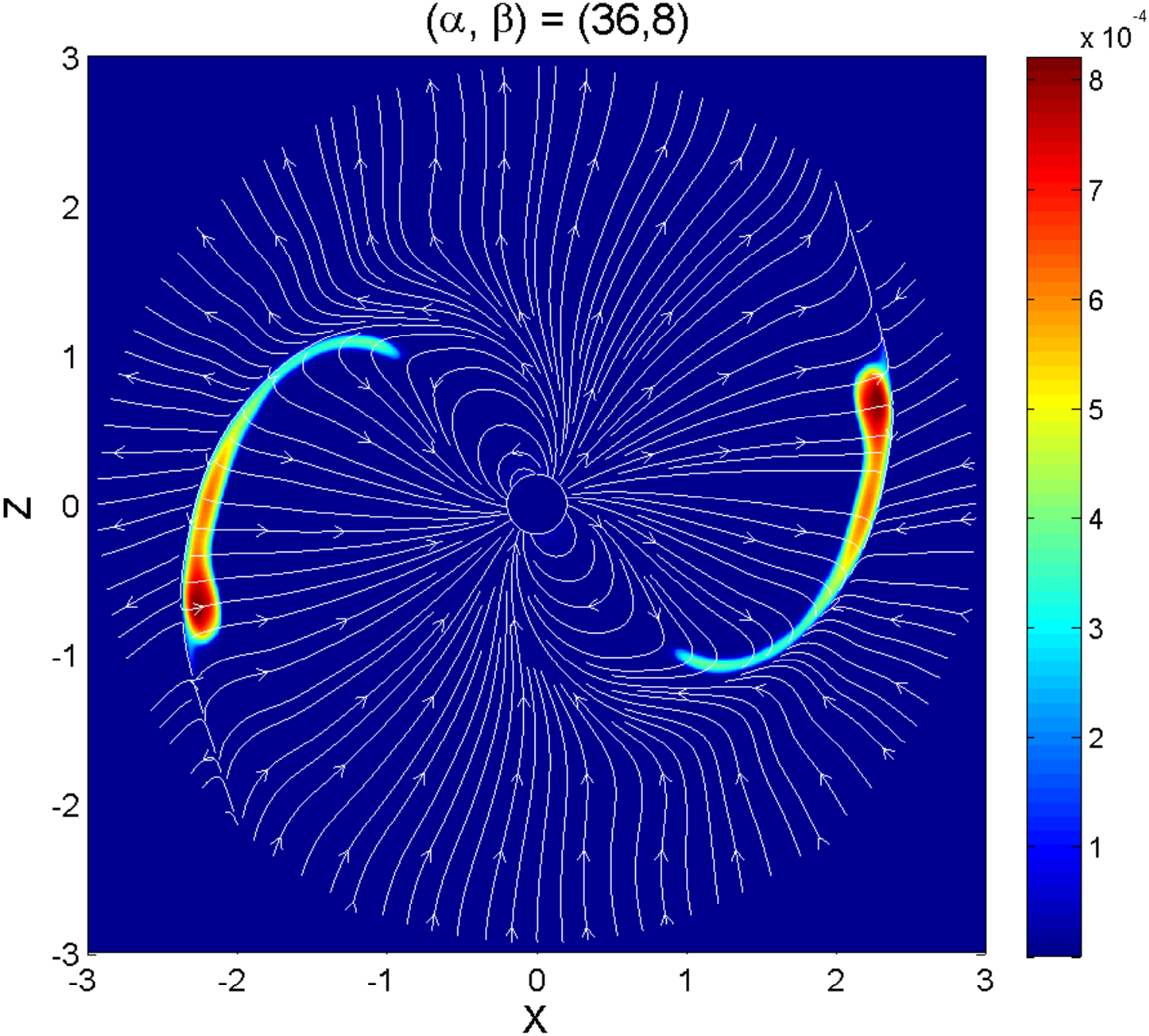} \qquad
\includegraphics[width=5.4 cm,height=5.5 cm]{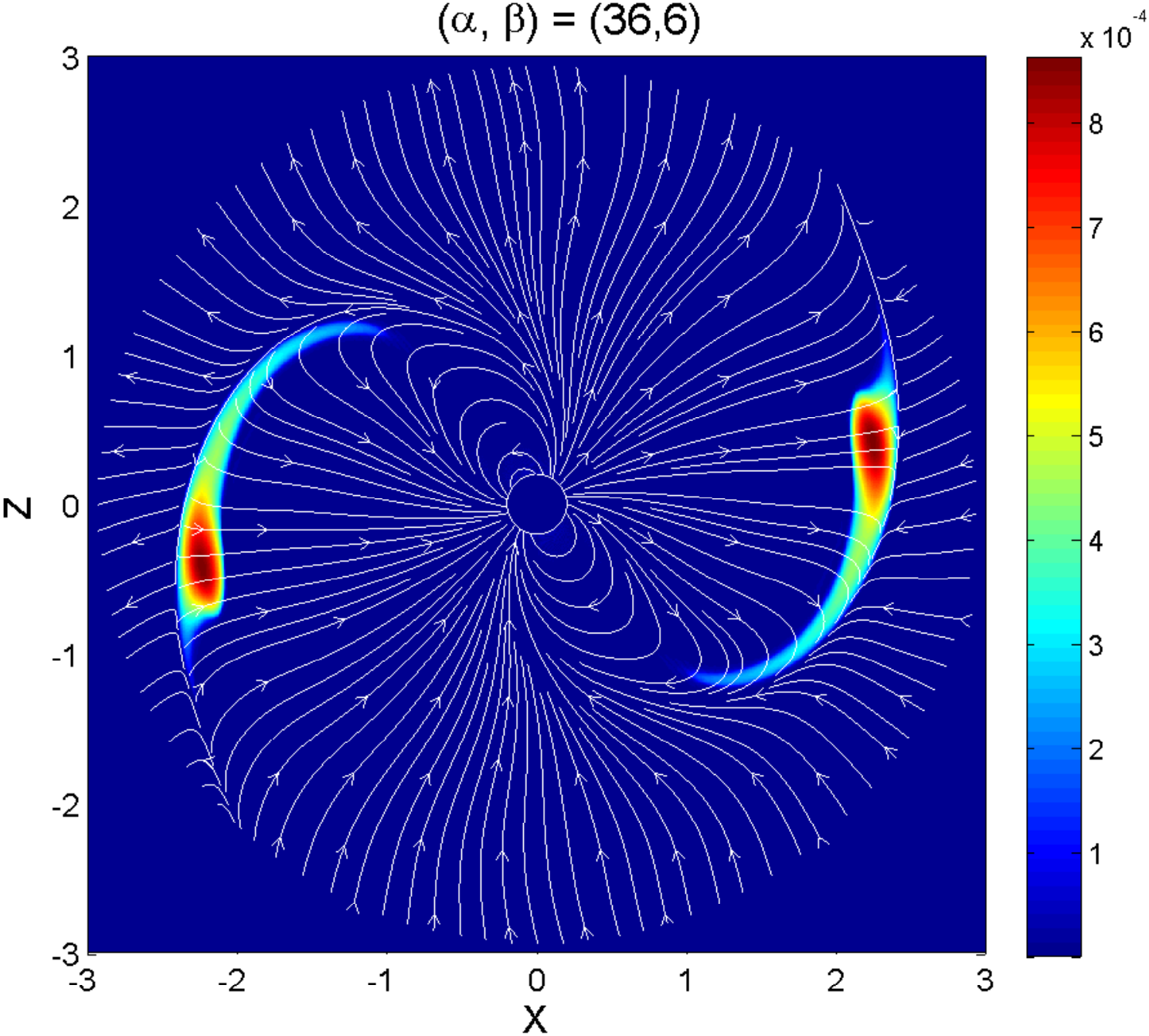} \qquad
\end{tabular}
\caption{Distribution of the magnetic field lines and the accelerating electric field $E_0$ in the $x$$-$$z$ plane for a $\chi=60^{\circ}$ dissipative rotator with the pair multiplicity $\kappa=1$ by using different filter parameters, where $E_0$ is the unit of the stellar surface magnetic field.\\ }
\label{Fig1}
\end{figure*}

\begin{figure}
\center
\begin{tabular}{cccccc}
\includegraphics[width=8.0cm,height=7.cm]{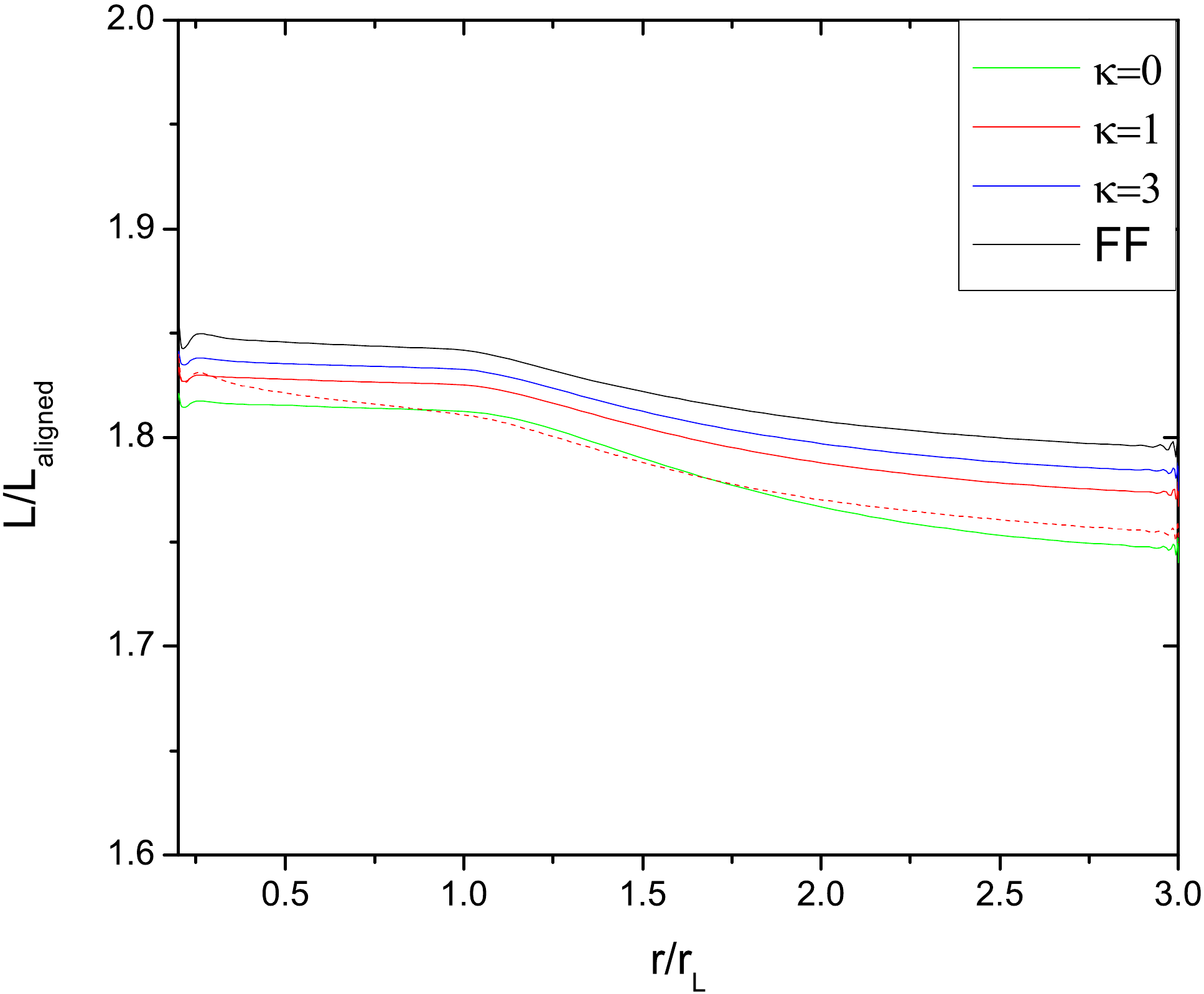}
\end{tabular}
\caption{The normalized Poynting flux $L/L_{\rm aligned}$ as a function of radius $r$ for a $\chi=60^{\circ}$ dissipative rotator  with different pair multiplicities $\kappa$. For comparison,
the normalized Poynting flux with the pair multiplicity $\kappa=1$ for the  low resolution of $N_r \times N_{\theta} \times N_{\phi}=129 \times 32 \times 64$ and the same optimized filtering parameter of  ($\alpha,\beta$)=(10,8) is also shown as the red dashed curve. }
\label{Fig2}
\end{figure}

\begin{figure*}
\center
\begin{tabular}{cccccccccccccc}
\includegraphics[width=5.4 cm,height=5.5 cm]{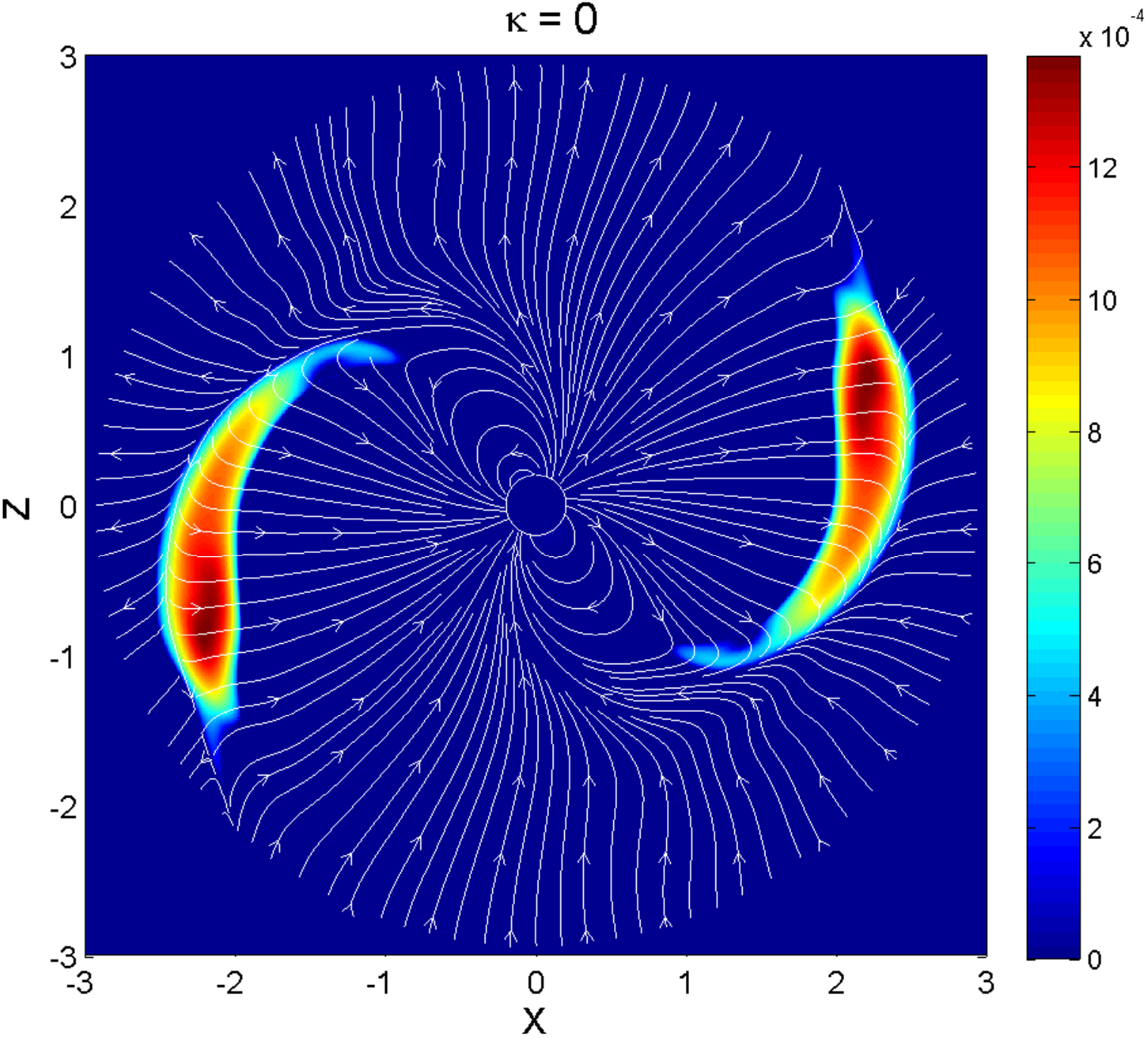} \qquad
\includegraphics[width=5.4 cm,height=5.5 cm]{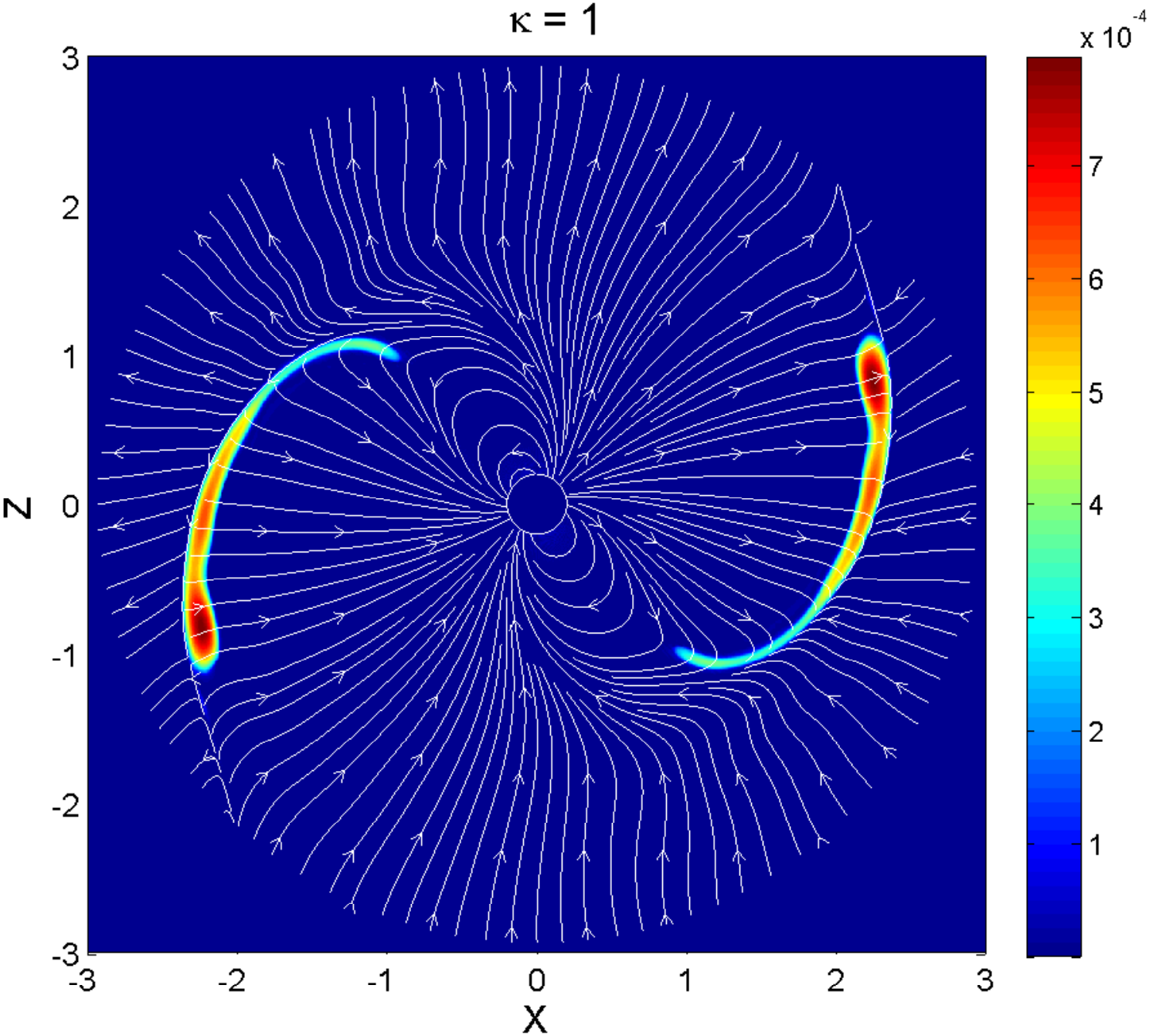} \qquad
\includegraphics[width=5.4 cm,height=5.5 cm]{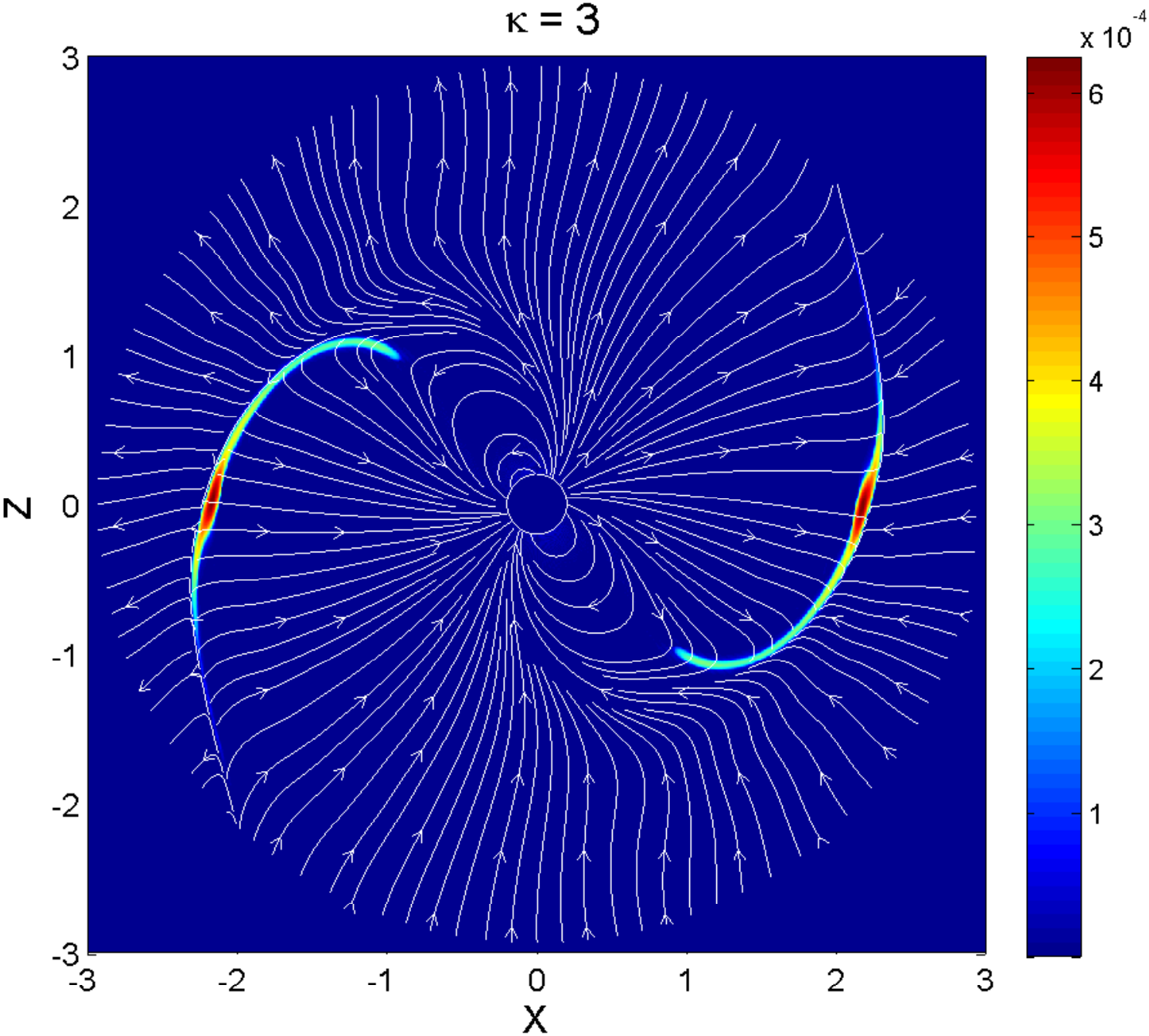} \qquad \\
\includegraphics[width=5.4 cm,height=5.5 cm]{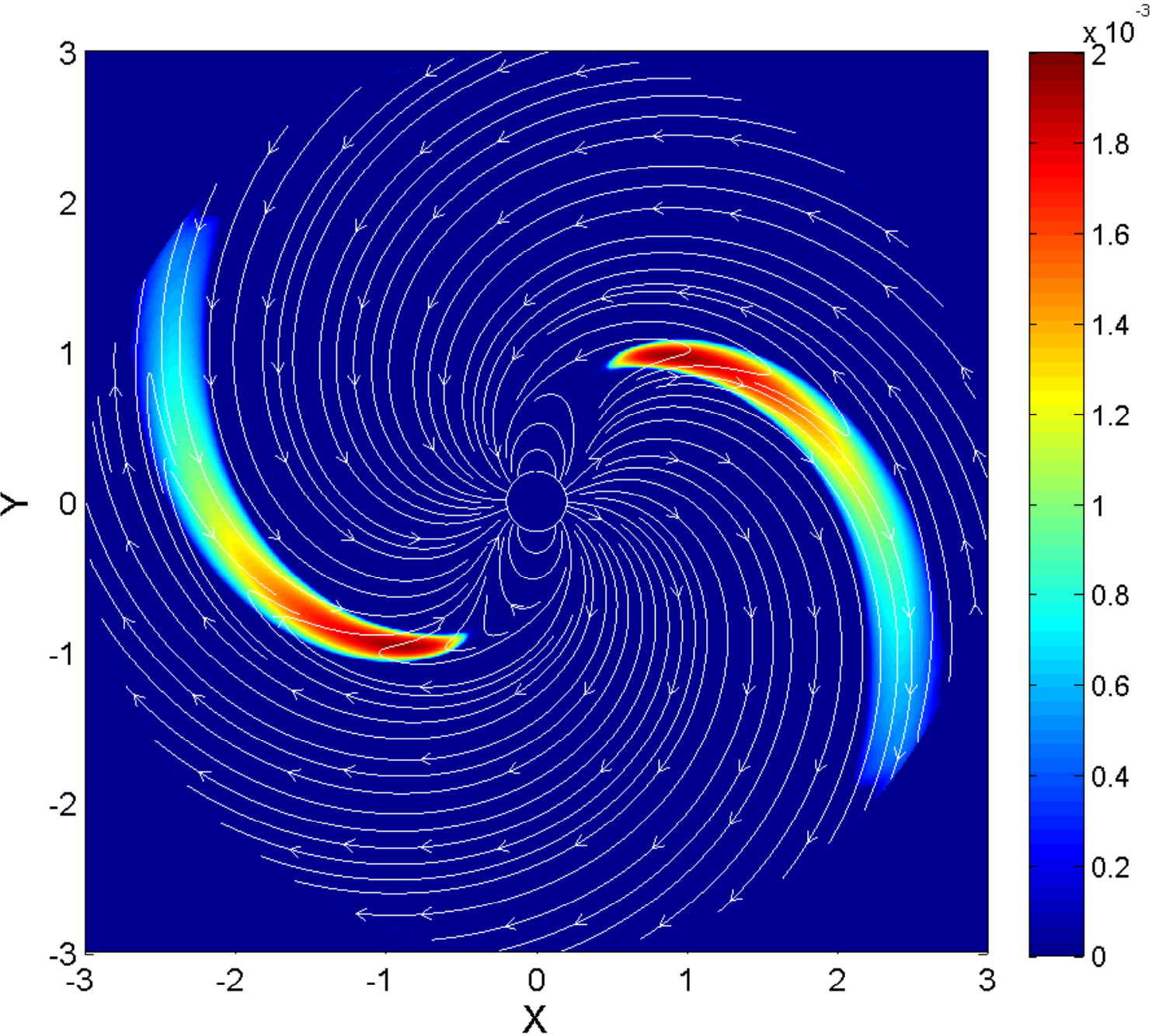} \qquad
\includegraphics[width=5.4 cm,height=5.5 cm]{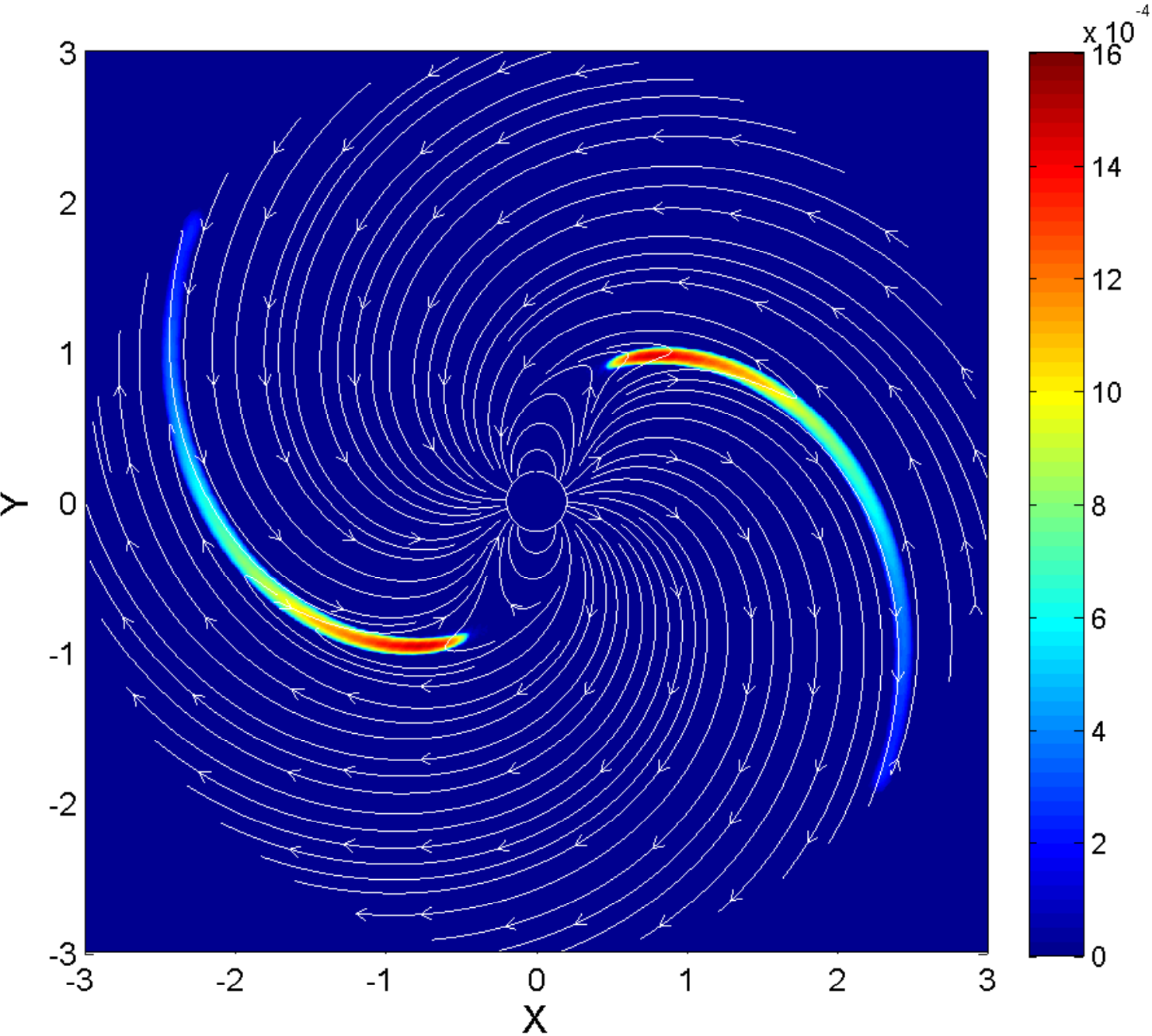} \qquad
\includegraphics[width=5.4 cm,height=5.5 cm]{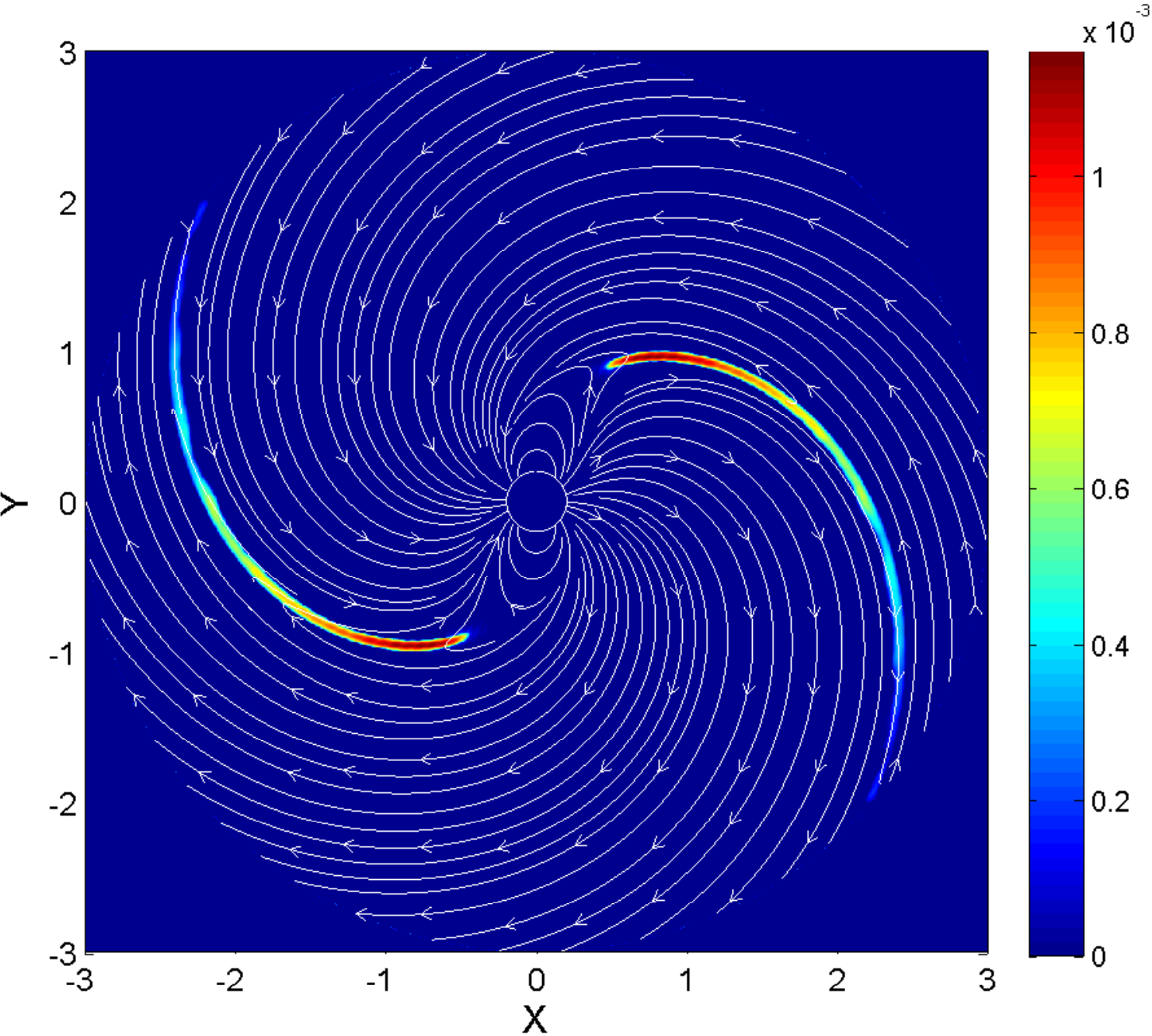} \qquad
\end{tabular}
\caption{Distribution of the magnetic field lines and the accelerating electric field $E_0$  for a $\chi=60^{\circ}$ dissipative rotator with the pair multiplicity $\kappa=\{0,1,3\}$ and the optimized filtering parameters ($\alpha,\beta$)=(10,8) in the $x$$-$$z$ plane (the top plane) and in the $x$$-$$y$ plane (the bottom plane), where $E_0$ is the unit of the stellar surface magnetic field.\\   }
\label{Fig3}
\end{figure*}

\begin{figure*}
\center
\begin{tabular}{cccccccccccccc}
\includegraphics[width=5.85 cm,height=5.5 cm]{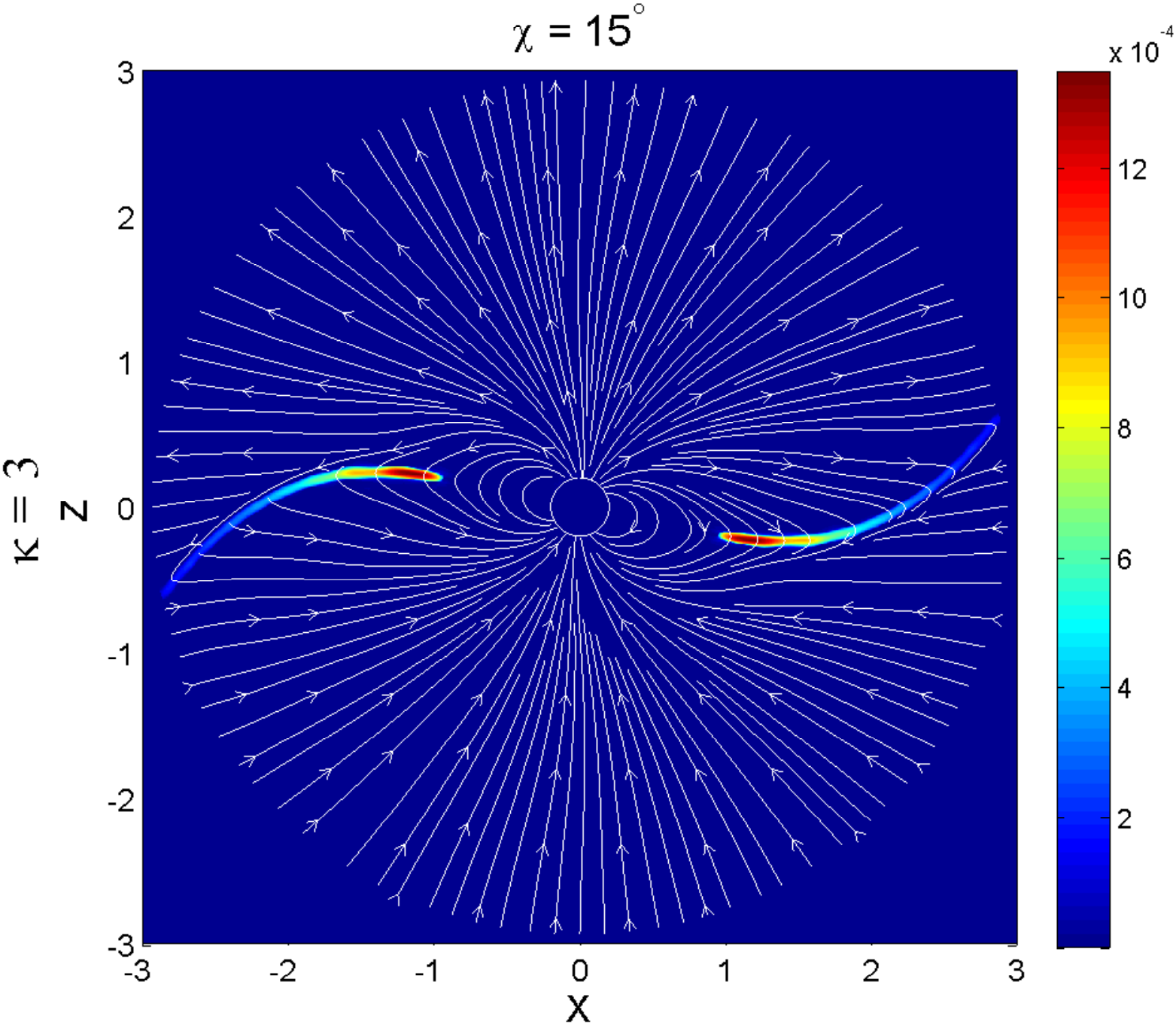} \qquad
\includegraphics[width=5.4 cm,height=5.5 cm]{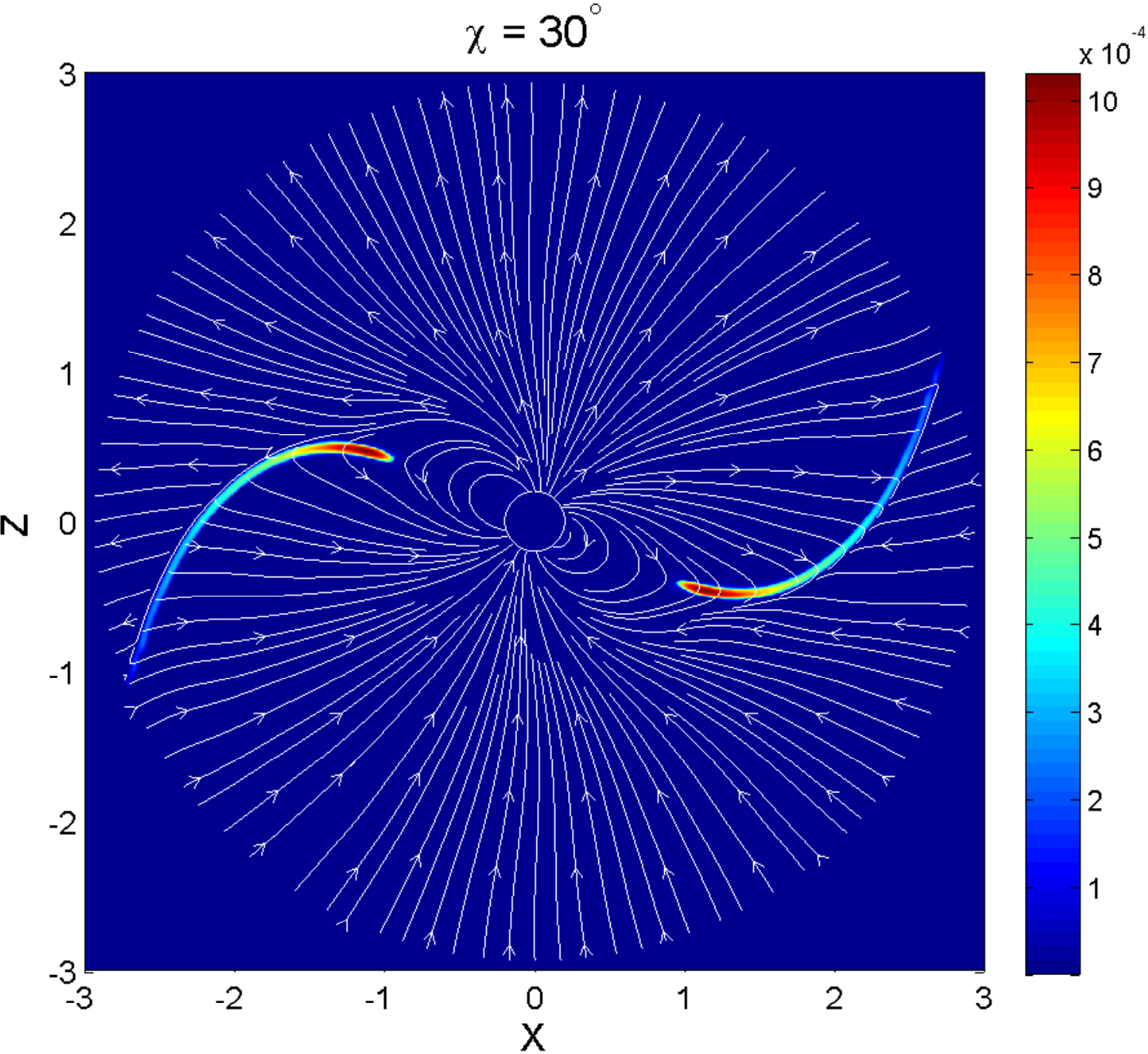} \qquad
\includegraphics[width=5.4 cm,height=5.5 cm]{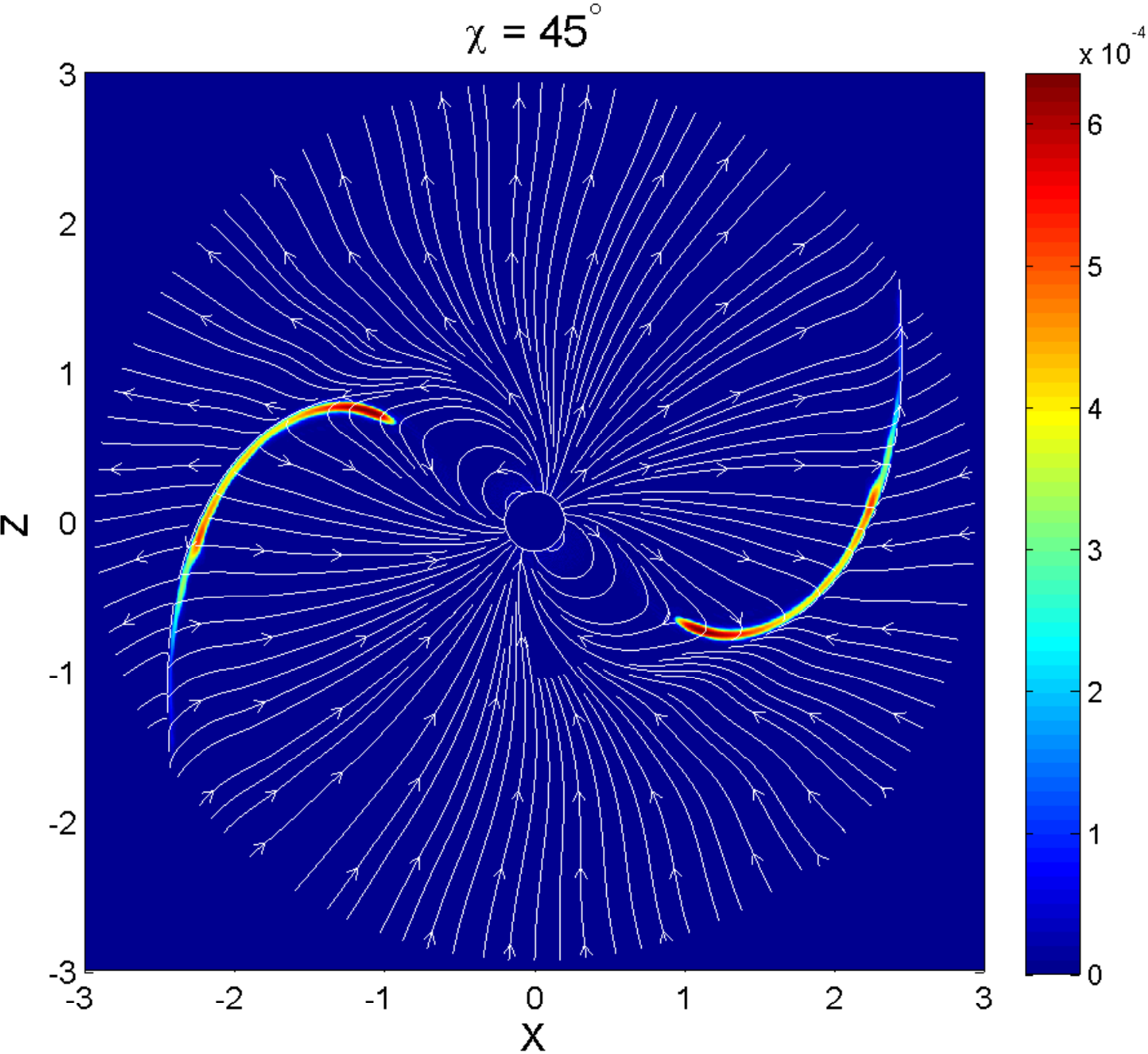} \qquad \\
\includegraphics[width=5.85 cm,height=5.5 cm]{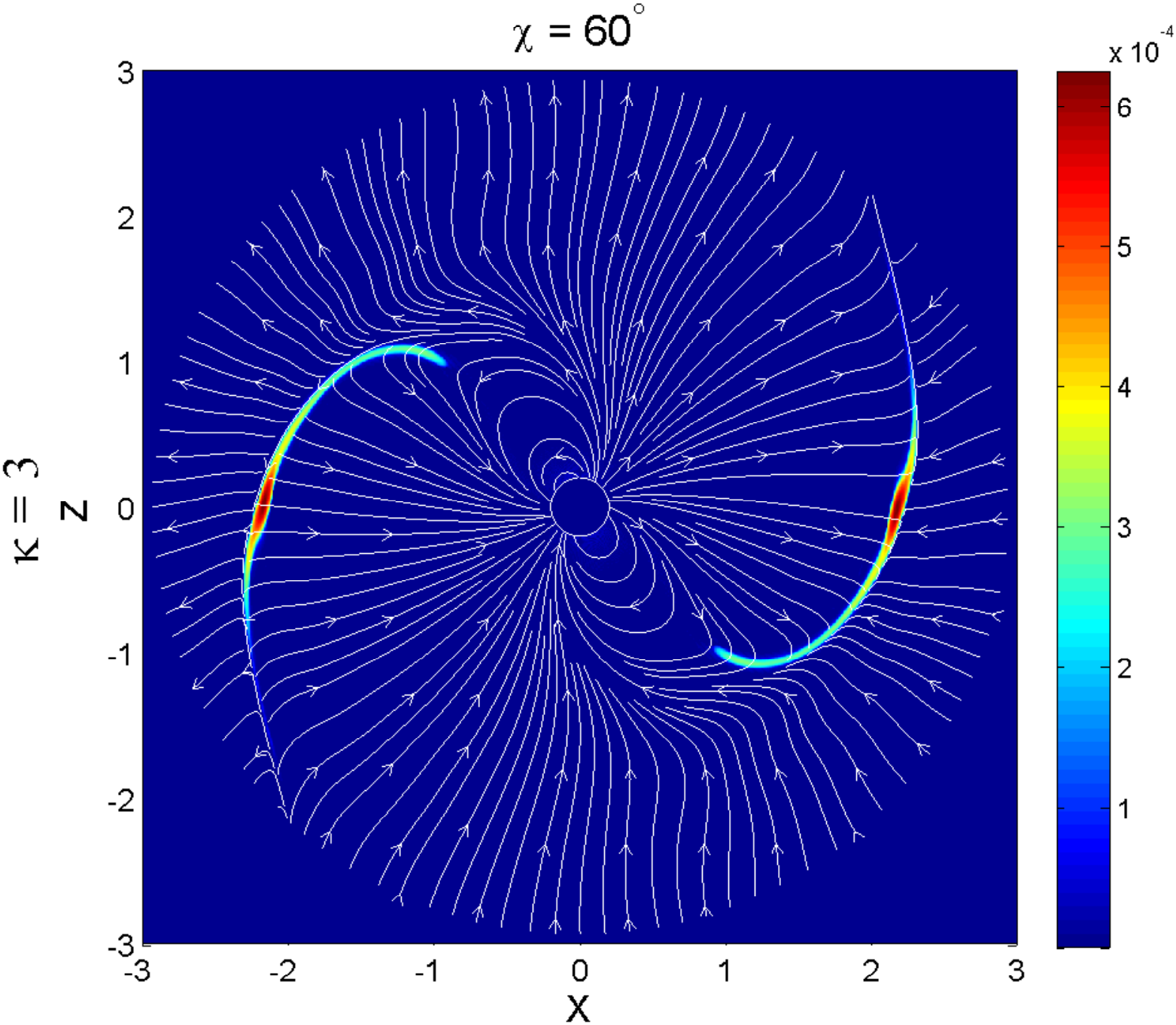} \qquad
\includegraphics[width=5.4 cm,height=5.5 cm]{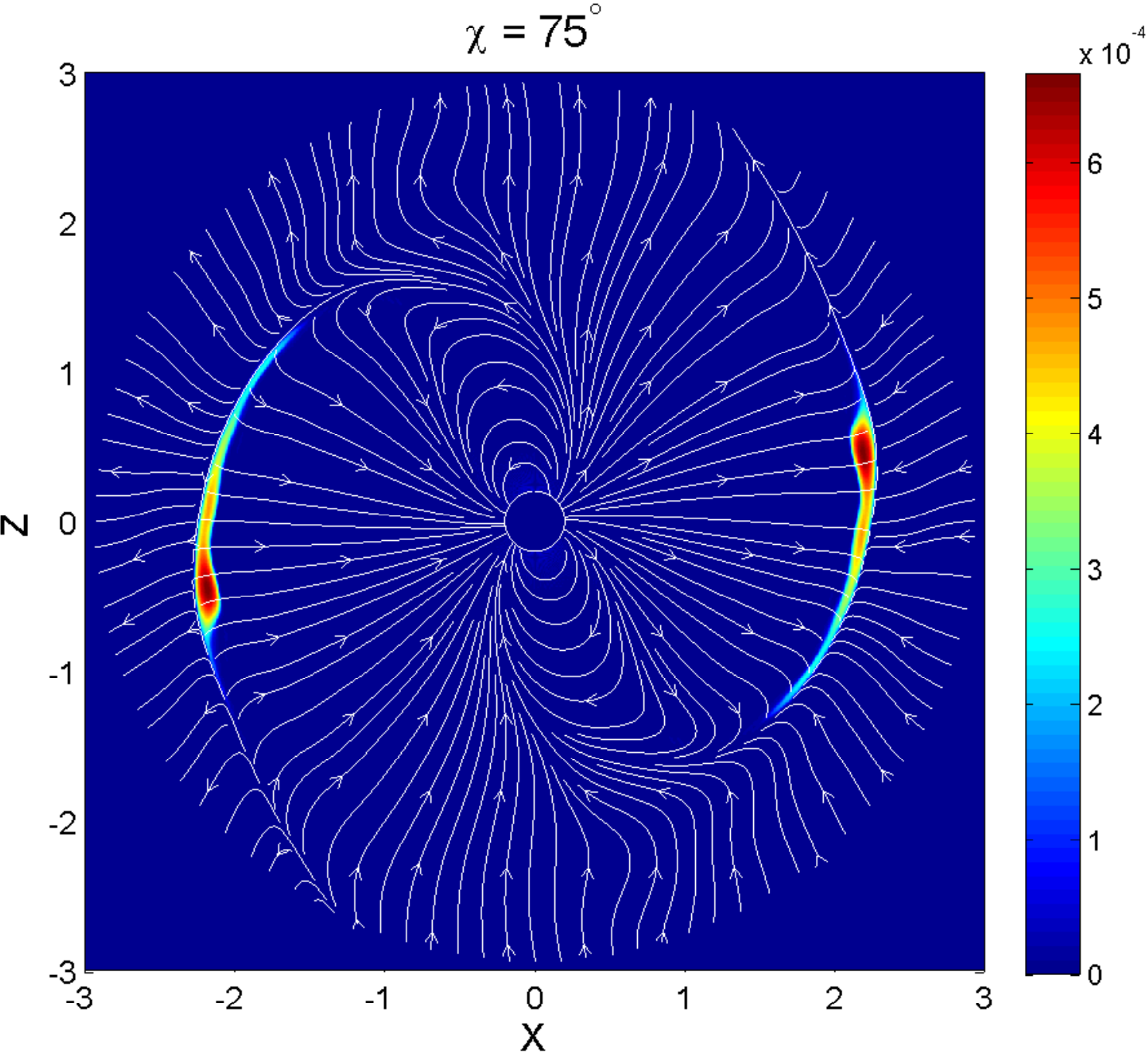} \qquad
\includegraphics[width=5.4 cm,height=5.5 cm]{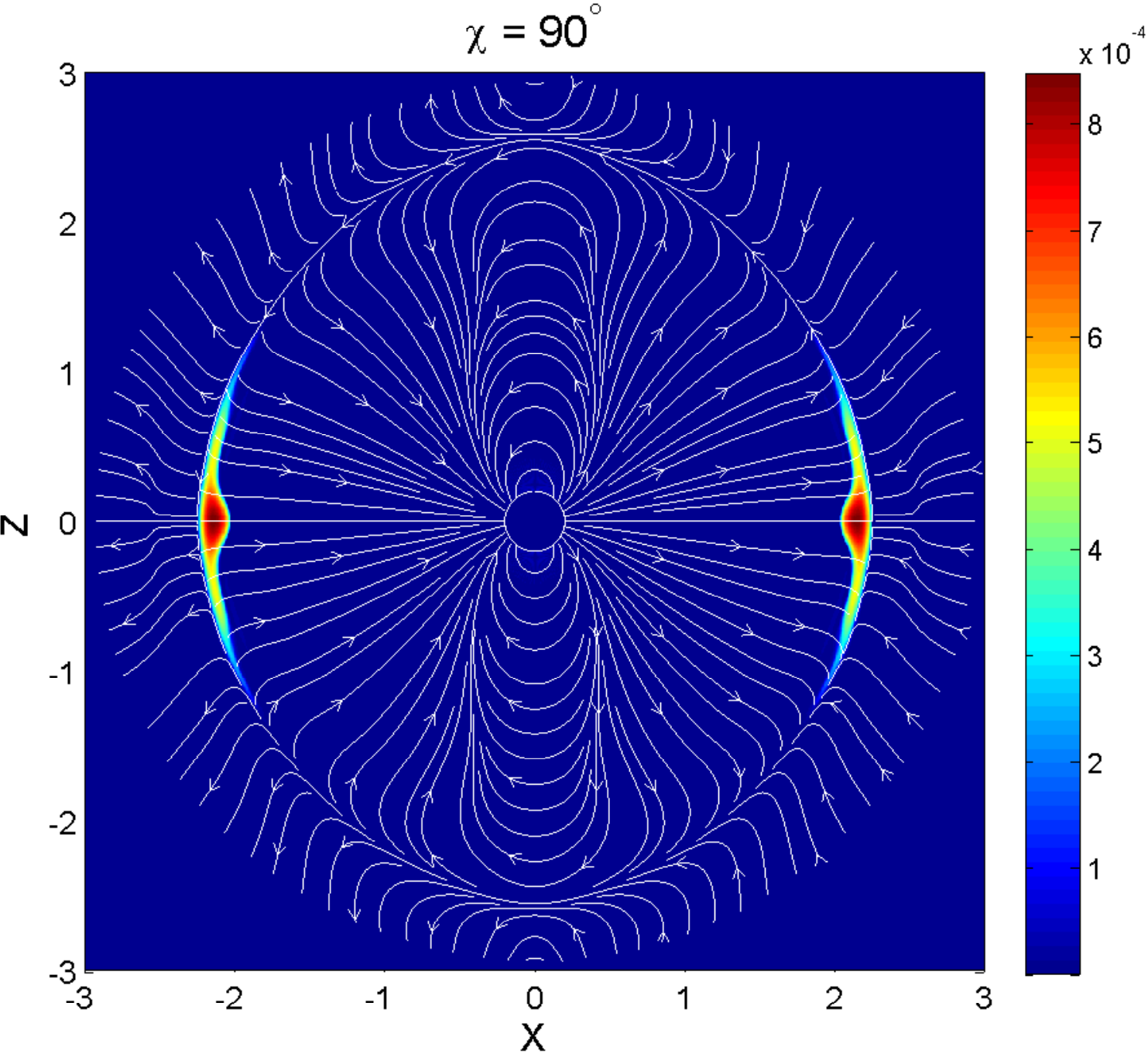} \qquad
\end{tabular}
\caption{Distributions of the magnetic field lines and the accelerating electric field $E_0$  for different magnetic inclination angles with the pair multiplicity $\kappa=3$ in the $x$$-$$z$ plane, where $E_0$ is the unit of the stellar surface magnetic field. \\ }
\label{Fig4}
\end{figure*}

\section{Aristotelian electrodynamics}
It is difficult to determine whether the magnetosphere has reached a steady state at the end of the simulation, especially  for the oblique rotator in which all the field lines remain
time-dependent in the observer frame. Therefore, it is useful to study the pulsar magnetosphere for stationary state in the co-moving frame, in which the magnetosphere relaxes to the time-independent solution where all the field lines remain time-independent.
The time-dependent Maxwell equations in the co-moving frame are given by \citep{mus05,pet20b}
\begin{eqnarray}
{\partial {\bf B}\over \partial t'}=-{\bf \nabla} \times ({\bf E}+{\bm V}_{\rm rot}\times{\bf B}) ,\\
{\partial  {\bf E}\over \partial t'}={\bf \nabla} \times ({\bf B}-{\bm V}_{\rm rot}\times{\bf E})-{\bf J}+{\bm V}_{\rm rot}\nabla\cdot{\bf E},\\
\nabla\cdot{\bf B}=0\;,\\
\nabla\cdot{\bf E}=\rho_{\rm e}\;,
\end{eqnarray}
where ${\bm V}_{\rm rot}={\bf \Omega } \times {\bf r}$ is the corotation velocity. $\rho_{\rm e}$ and ${\bf J}$ is the charge density and the current density, respectively. It is noted that {\bf E} and {\bf B} are still defined in the observed frame. The pulsar magnetosphere can be computed by implementing a prescription for the current density ${\bf{J}}$.

The pulsar magnetosphere cannot be surrounded by vacuum, because the rotating vacuum solution produces an accelerating electric field which is able to extract the particles from  the stellar surface  and fill the magnetosphere. Therefore, the realistic pulsar magnetospheres require the presence of plasma and have some dissipation regions to produce the particle acceleration  and the pulsed radiation. The emitted photon has
a back-reaction onto the particle motion in a direction opposite to its motion. This process can be easily treated by assuming a stationary balance between the particle acceleration and radiation.
which is called Aristotelian electrodynamics. The  current density in the AE magnetosphere can be defined by introducing the pair multiplicity $\kappa$ as \citep{cao20}
\begin{eqnarray}
{\bf J }=  \rho_e  \frac {{\bf E} \times {\bf B}}{B^2+E^2_{0}}+ (1+\kappa)\left|\rho_e\right| \frac{ (B_0{\bf {B}}+E_0{\bf {E}}) }{ B^2+E^2_{0}}.
\label{Eq6}
\end{eqnarray}
where $B_0$ and $E_0$ are the magnetic and electric field in the frame in which ${\bf E}$ and ${\bf B}$ are parallel. The quantities $B_0$ and $E_0$ is given by the
relations
\begin{eqnarray}
B^2_{0}-E^2_{0}&=&{\bf B}^2-{\bf E}^2,\\
 E_{0}B_{0}&=&{\bf E}\cdot {\bf B},
\end{eqnarray}
with $E_{0}\geq0$.

The force-free approximation  satisfies the force-free condition ${\bf E}\cdot {\bf B}=0$ and requires the  condition $E \leq B$ in the whole magnetosphere. The current sheet is captured by enforcing the condition $E = B$ in the regions where $E>B$. Therefore, the force-free approximation can not produce any dissipative regions where $E > B$. However, the AE model can allow for dissipation in some regions where $E > B$. It is well known that the pulsar magnetosphere should allow for a local dissipation to produce the pulsed emission. The recent numerical simulations shown that the current sheet is a promising site for  particle acceleration and high-energy radiation in the magnetosphere. To explore the dissipation  where $E > B$, we use the force-free description where $E \leq B$ and the AE  description where $E > B$. It is expected that our model will  produce the dissipative region with $E > B$ outside the LC.

\begin{figure}
\begin{tabular}{cccccc}
\includegraphics[width=7 cm,height=6.25 cm]{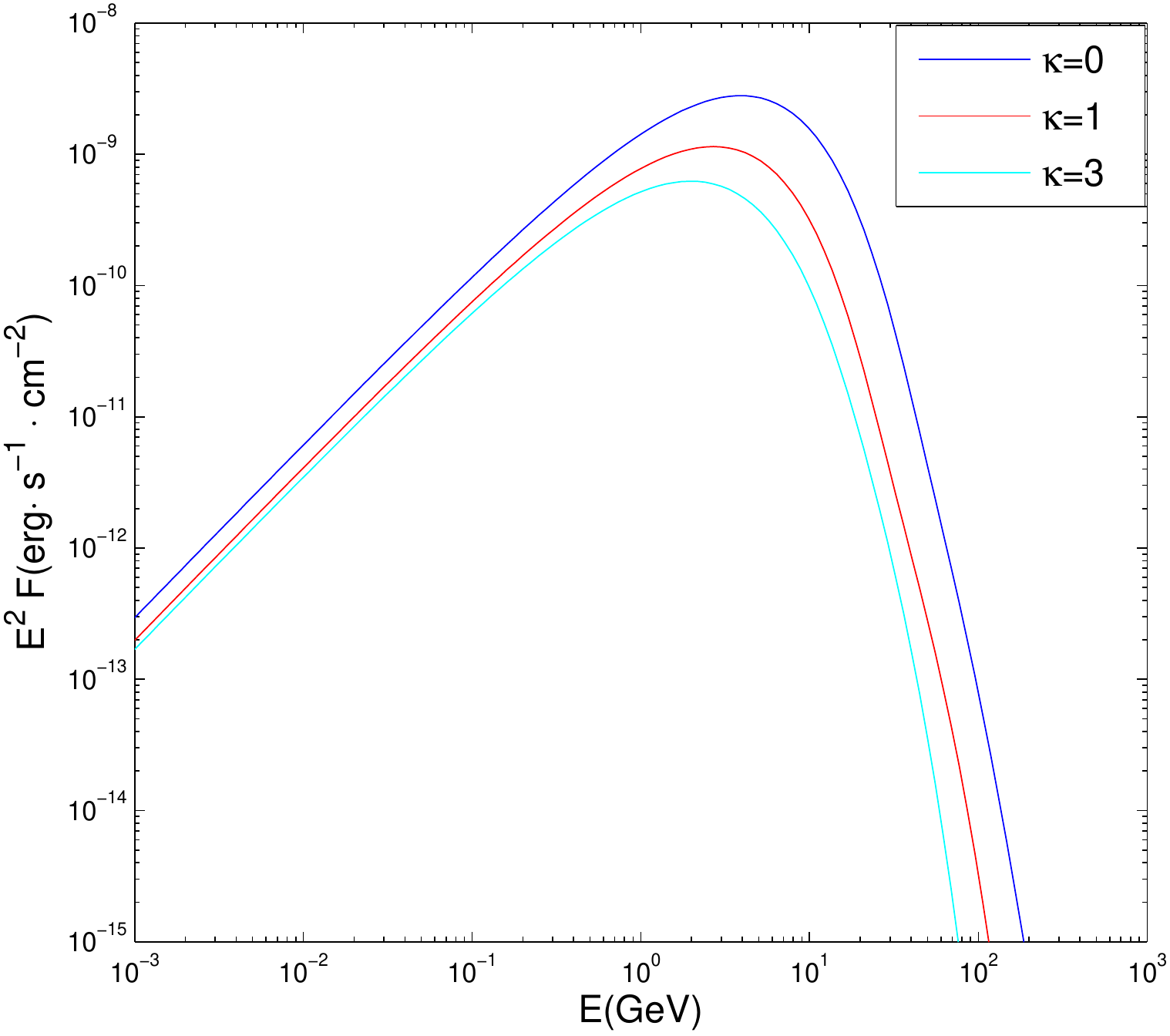}
\end{tabular}
\caption{The curvature radiation spectra for a  $60^\circ$ magnetosphere with the pair multiplicities $\kappa=\{0,1,3\}$. }
\label{Fig5}
\end{figure}

\section{result}
\subsection{Magnetospheric Structure}
The time-dependent Maxwell equations are solved by a spectral algorithm in the co-moving frame with the combined FFE and AE description. The electromagnetic field are discretized by a  set of the spectral collocation points  in spherical coordinates ($r$,\,$\theta$,\,$\phi$). The radial components of the electromagnetic field are expanded onto the Chebyshev polynomial, and the angle components are expanded onto the vector spherical harmonic expansion. We improve the time-integration described in \citet{cao20} by using a combined three-order Runge-Kutta and Adam-Bashforth method. An exponential filter with $\sigma(\eta) = \textrm{exp}(-\alpha\,\eta^\beta)$ in all directions is used to smooth the electromagnetic field at each time step. The divergenceless condition on the magnetic field is  analytically enforced  by  a projection method. The simulation is initialized with  a  dipole  magnetic field  and a zero electric field outside the star. We enforce the inner boundary condition  at the stellar surface with a rotating electric field ${\bf {E}} = -( {\bf \Omega } \times {\bf r} ) \times {\bf B}/c$. We use a non-reflecting boundary condition to prevent  reflection from the outer boundary. The simulation domain is set  to  $r\in (0.2 - 3)$ $r_{\rm L}$. A high resolution with $N_r \times N_{\theta} \times N_{\phi}=129 \times 64 \times 128$ is used to catch the current sheet in all simulations. We performed several simulations for magnetic inclination angle $\chi= \{0^\circ,15^\circ,30^\circ,45^\circ,60^\circ,75^\circ,90^\circ\}$ with the pair multiplicity $\kappa=\{0,1,3\}$. The system  is evolved for several rotational periods to ensure that a  steady solution has been reached.


\begin{figure*}
\center
\begin{tabular}{cccccccccccccc}
\includegraphics[width=5.7 cm,height=5.5 cm]{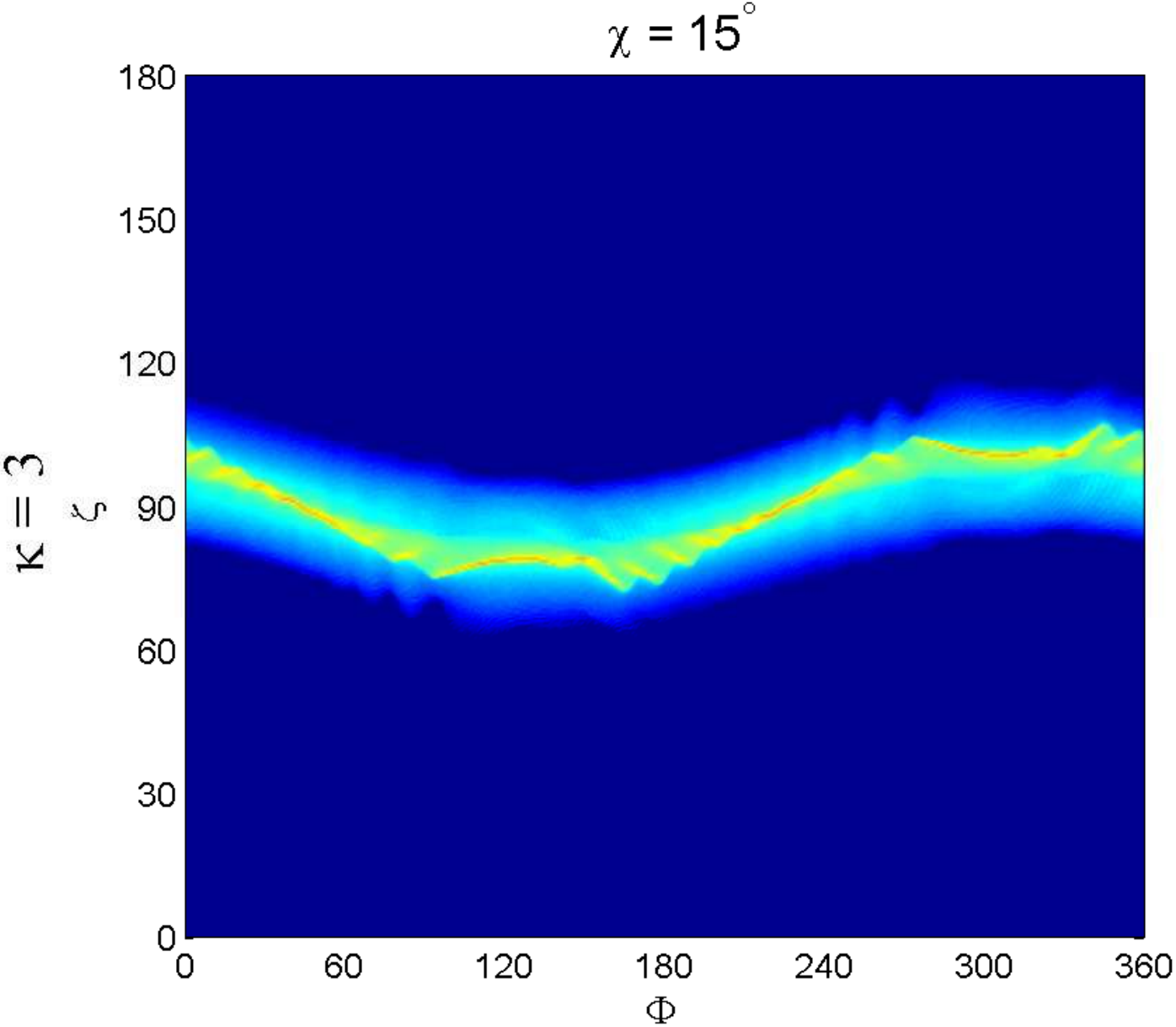} \qquad
\includegraphics[width=5.3 cm,height=5.5 cm]{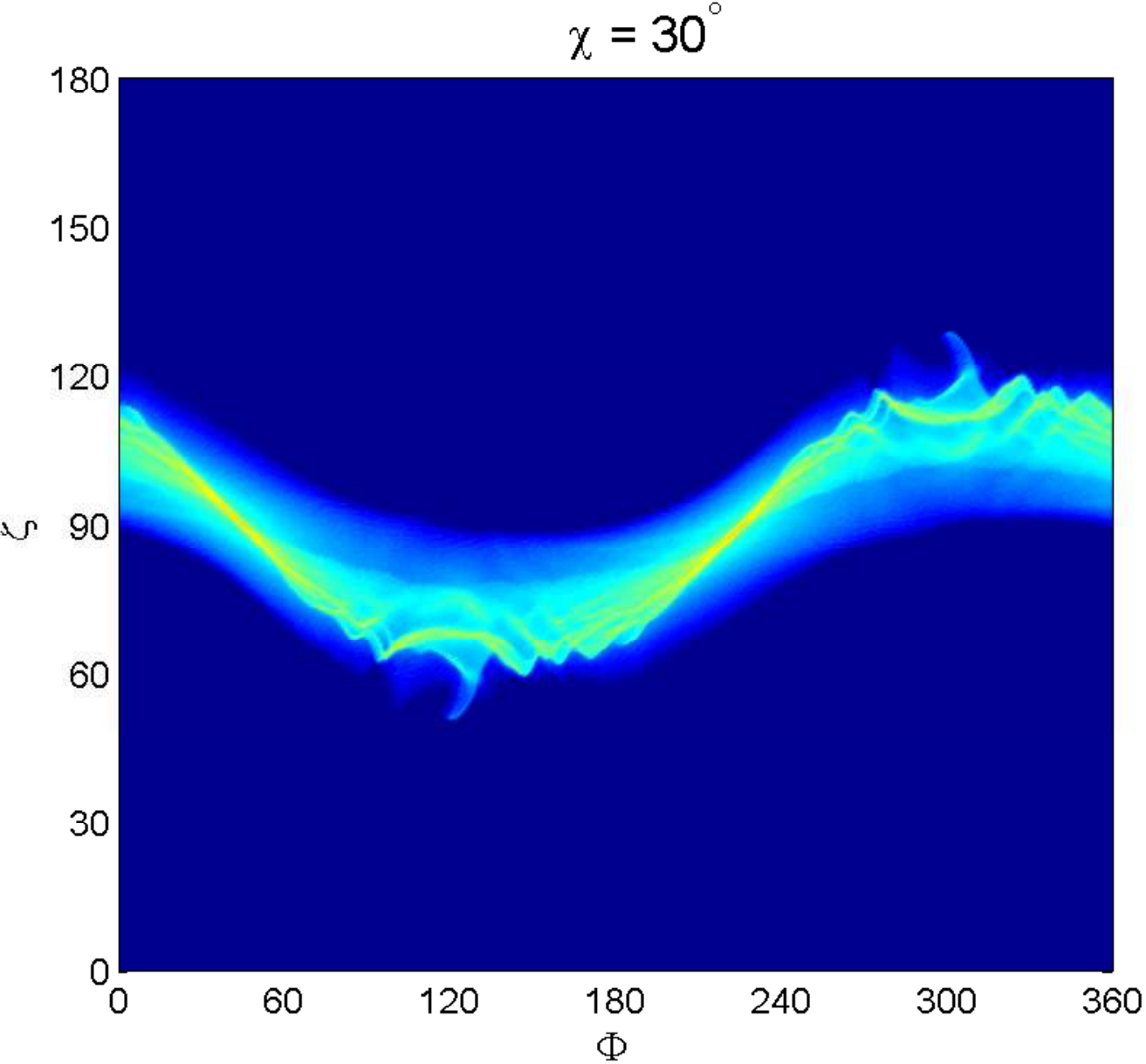} \qquad
\includegraphics[width=5.6 cm,height=5.5 cm]{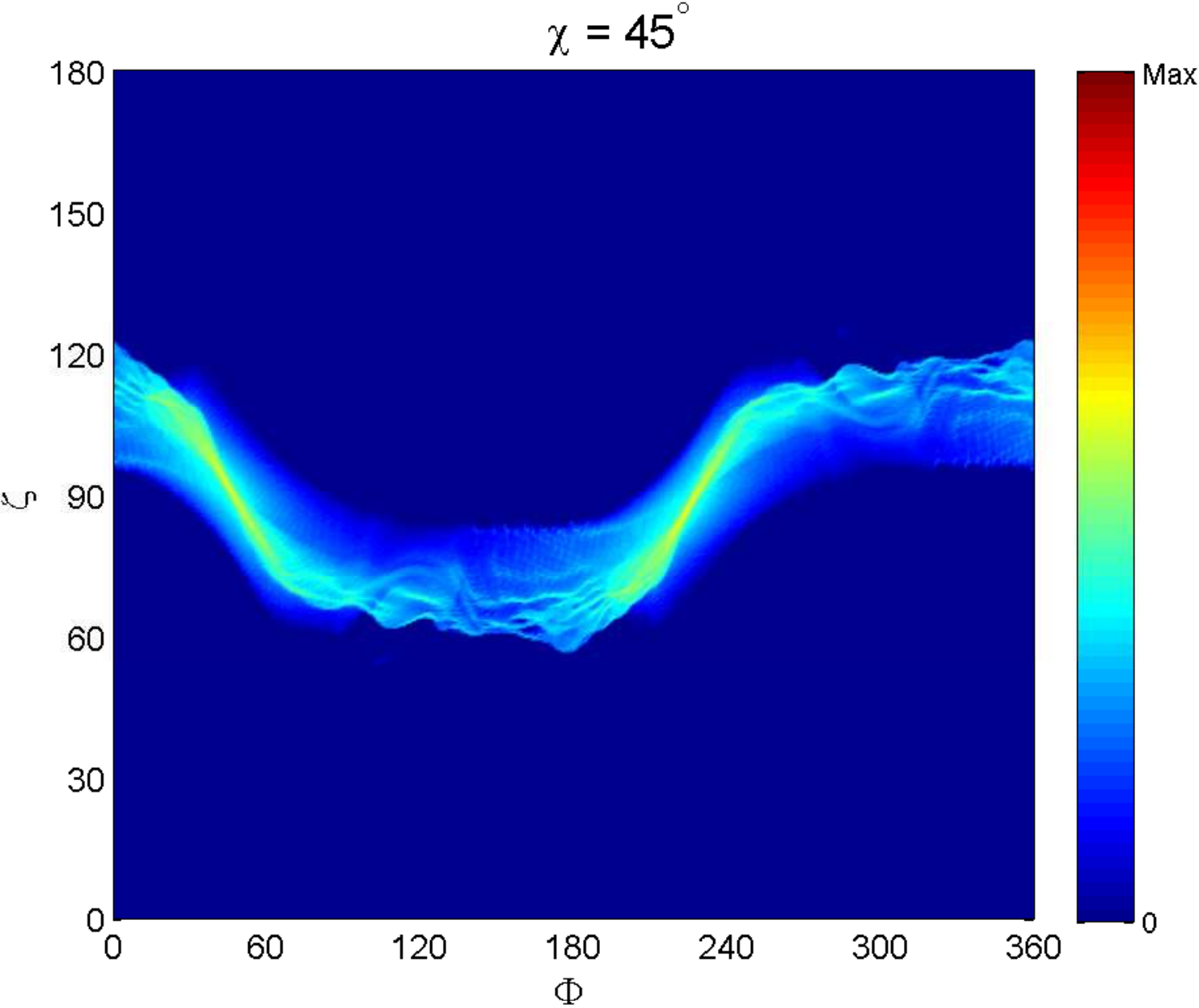} \qquad \\
\includegraphics[width=5.25 cm,height=5.35 cm]{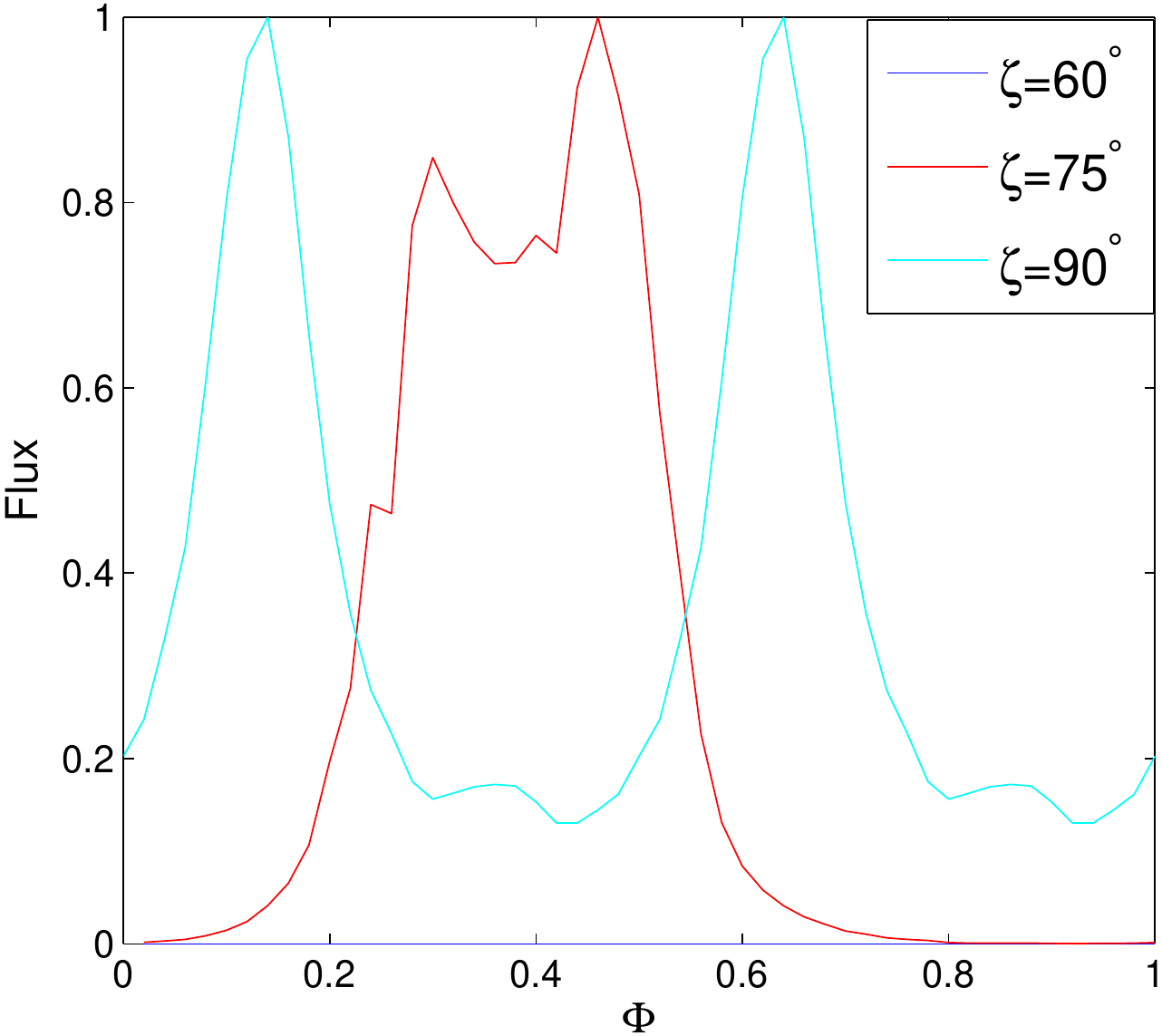} \qquad
\includegraphics[width=5.25 cm,height=5.35 cm]{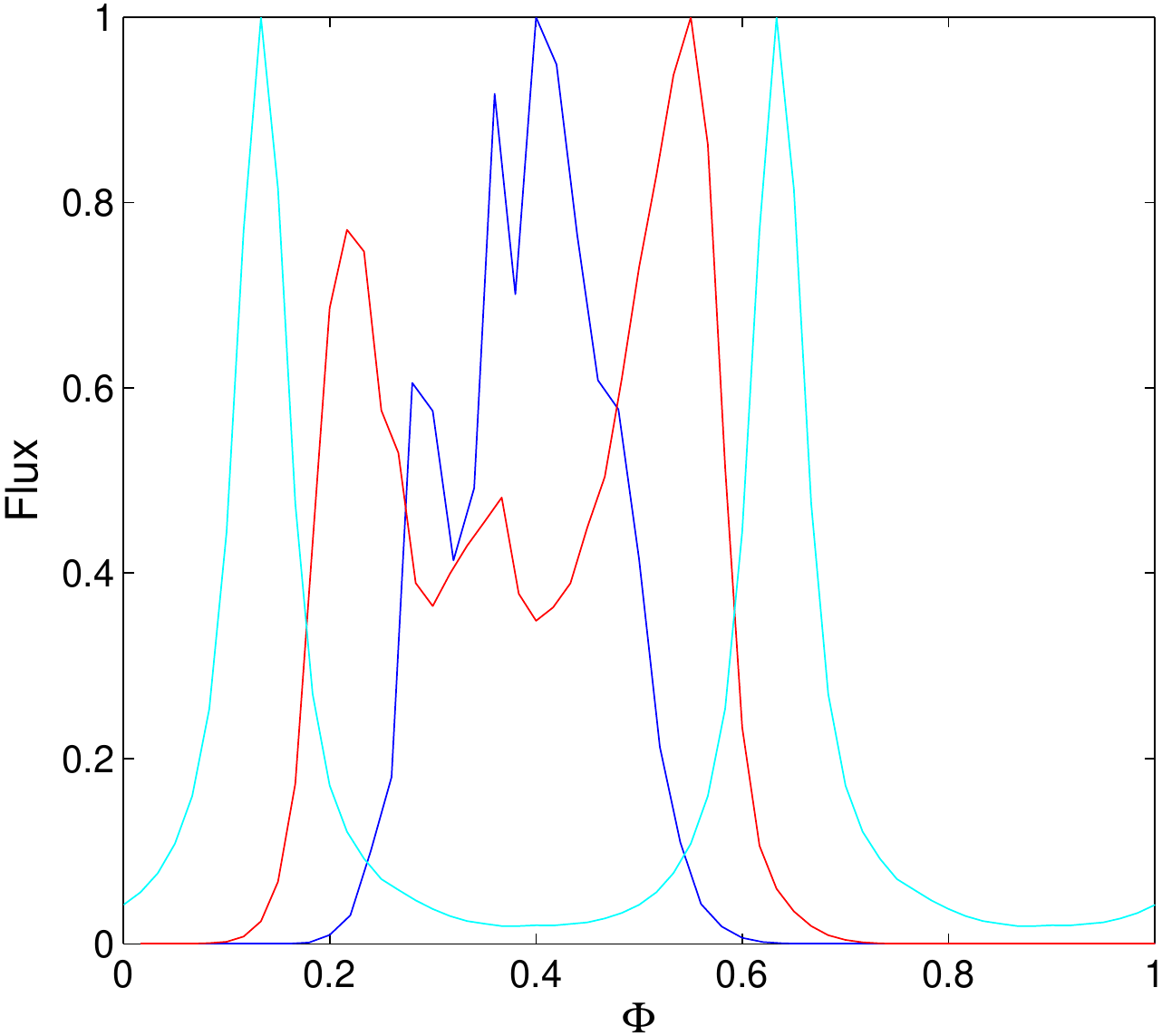} \qquad
\includegraphics[width=5.25 cm,height=5.35 cm]{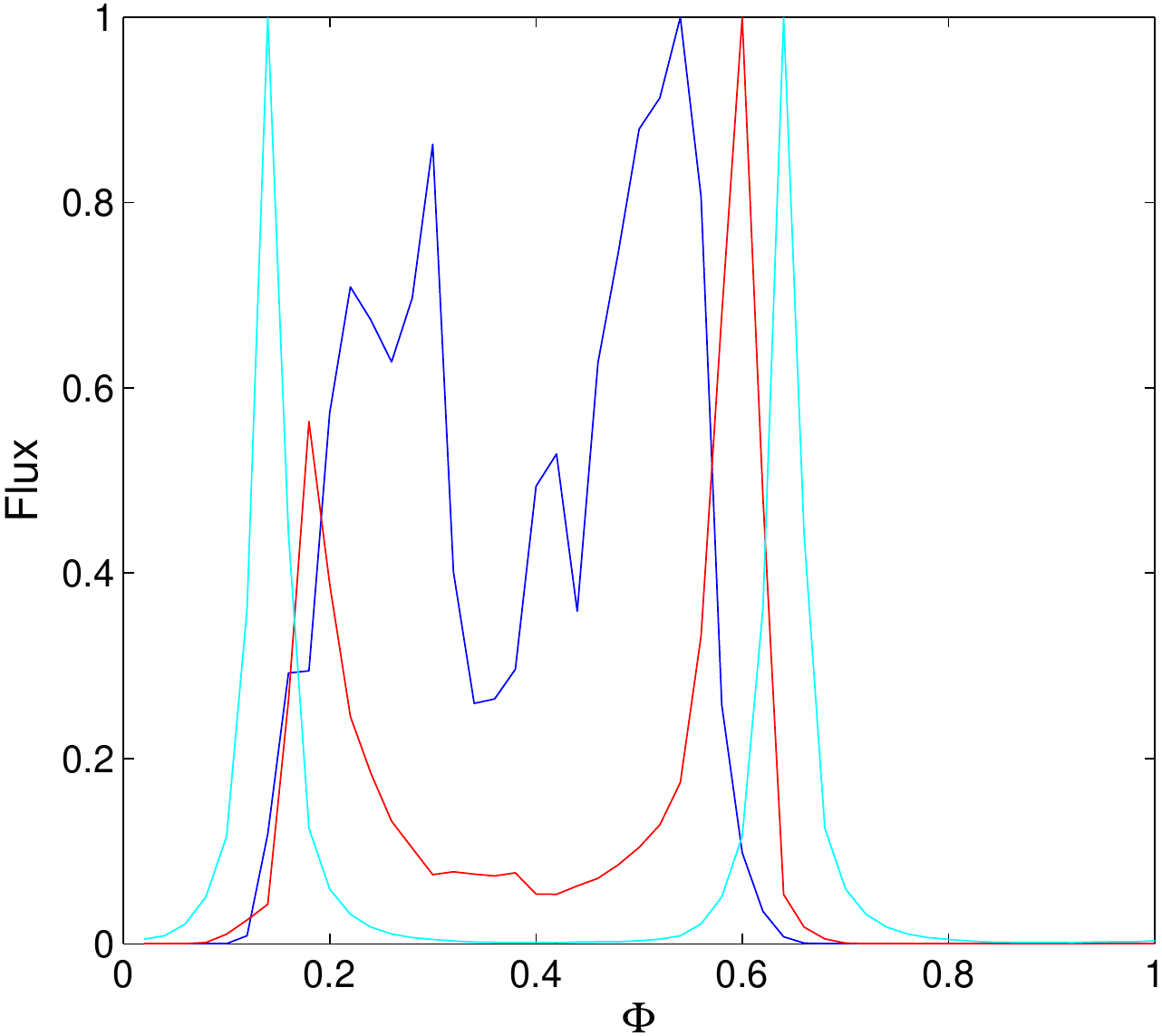} \qquad\\
\includegraphics[width=5.7 cm,height=5.5 cm]{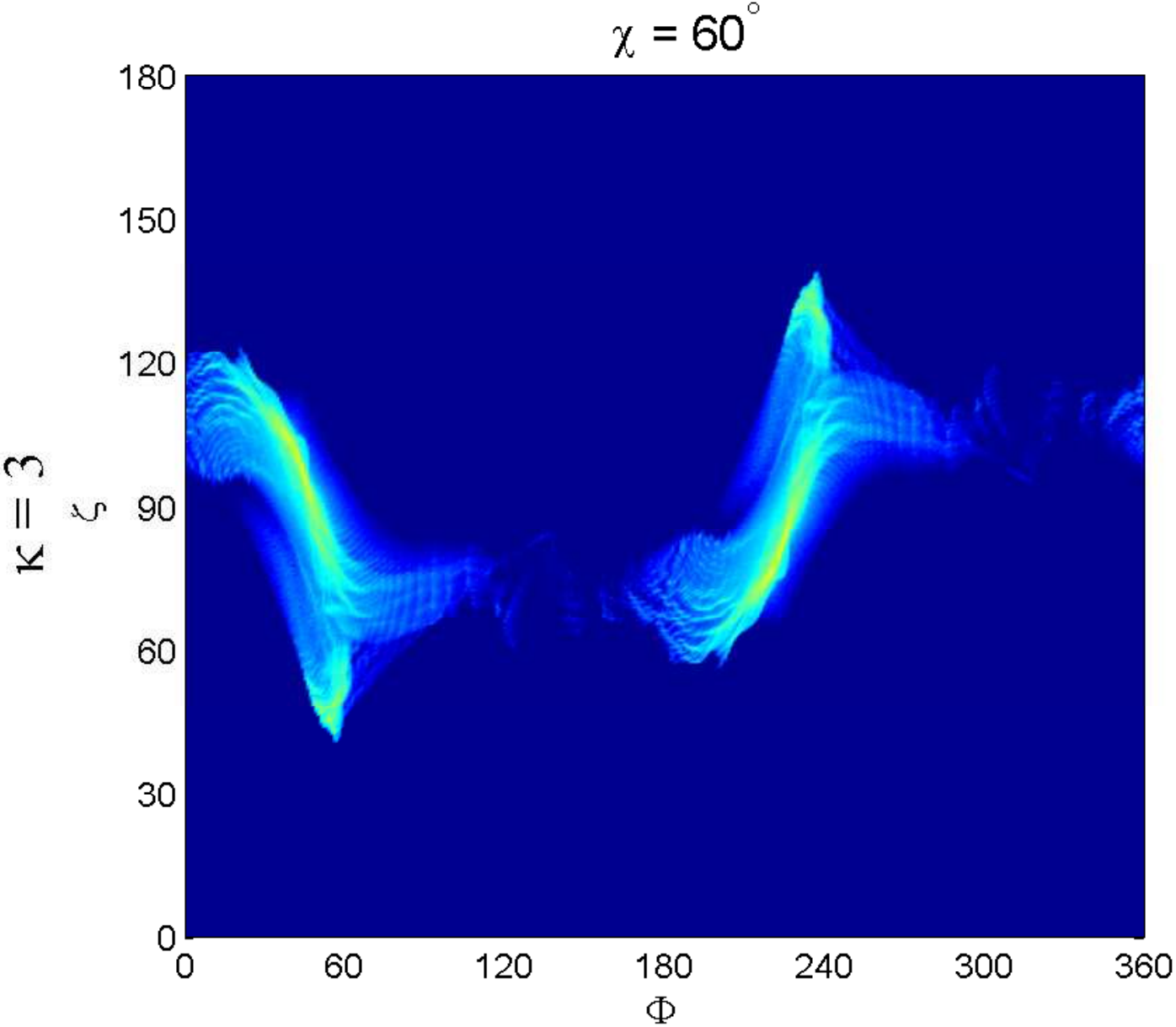} \qquad
\includegraphics[width=5.3 cm,height=5.5 cm]{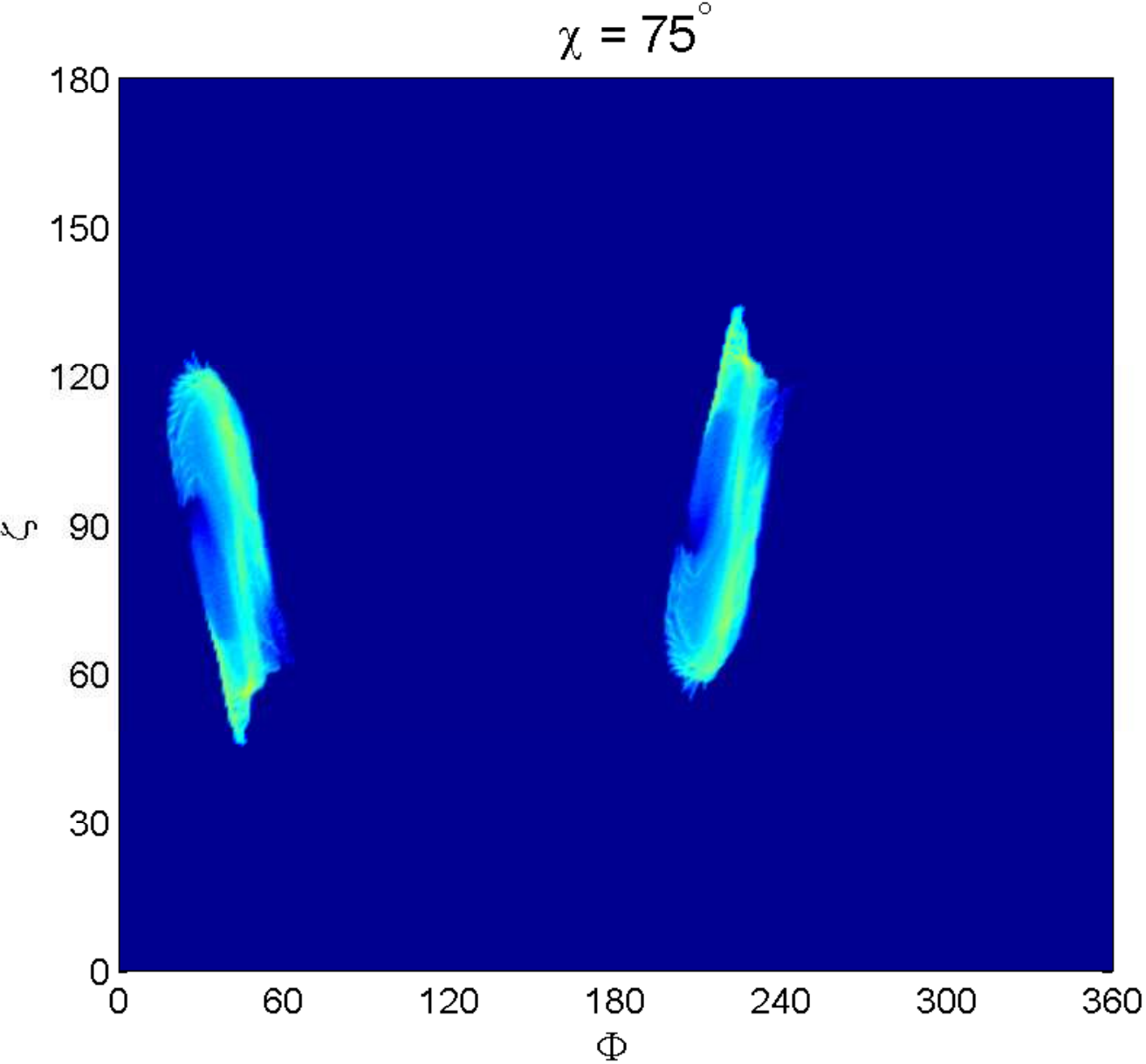} \qquad
\includegraphics[width=5.6 cm,height=5.5 cm]{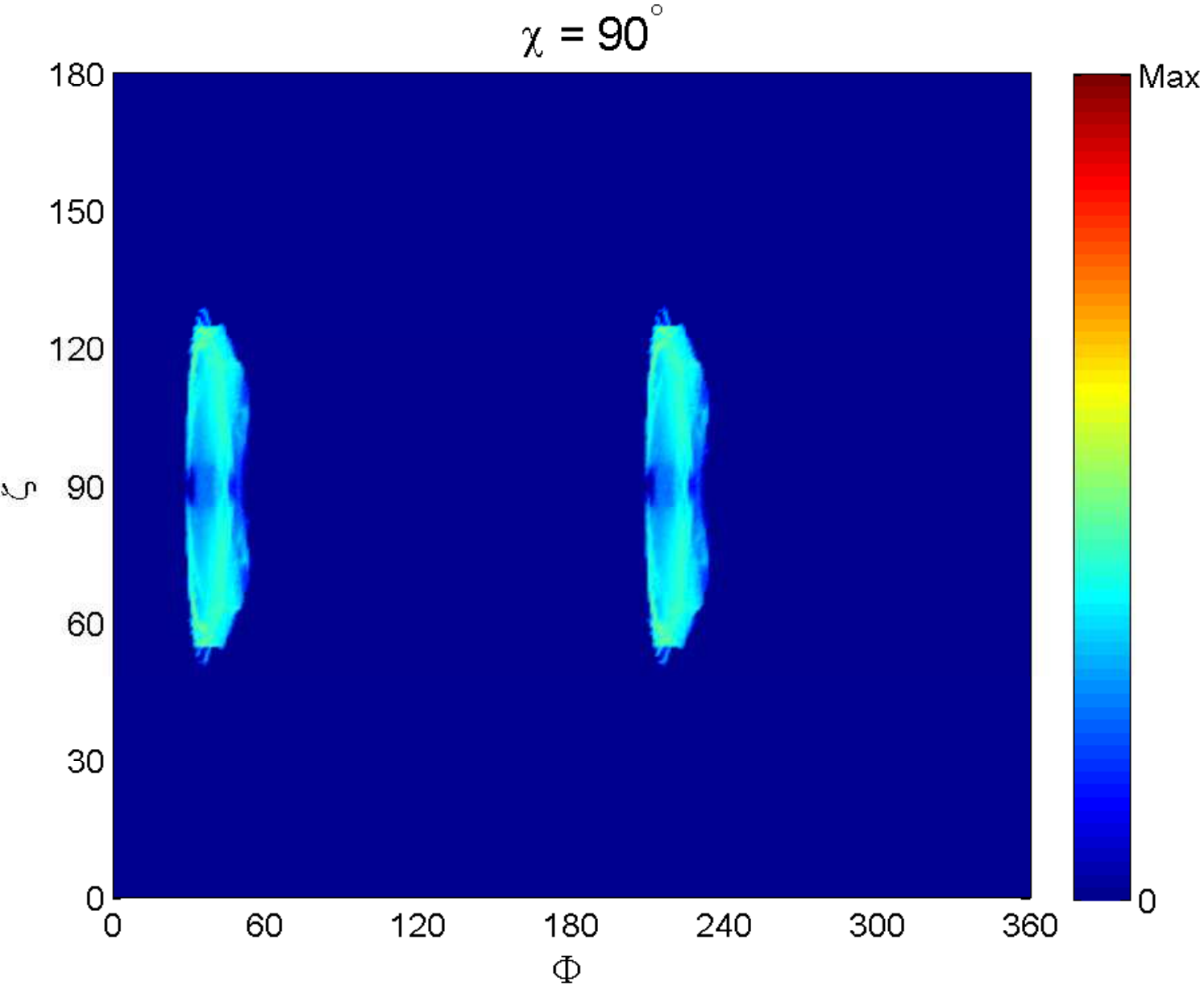} \qquad \\
\includegraphics[width=5.25 cm,height=5.35 cm]{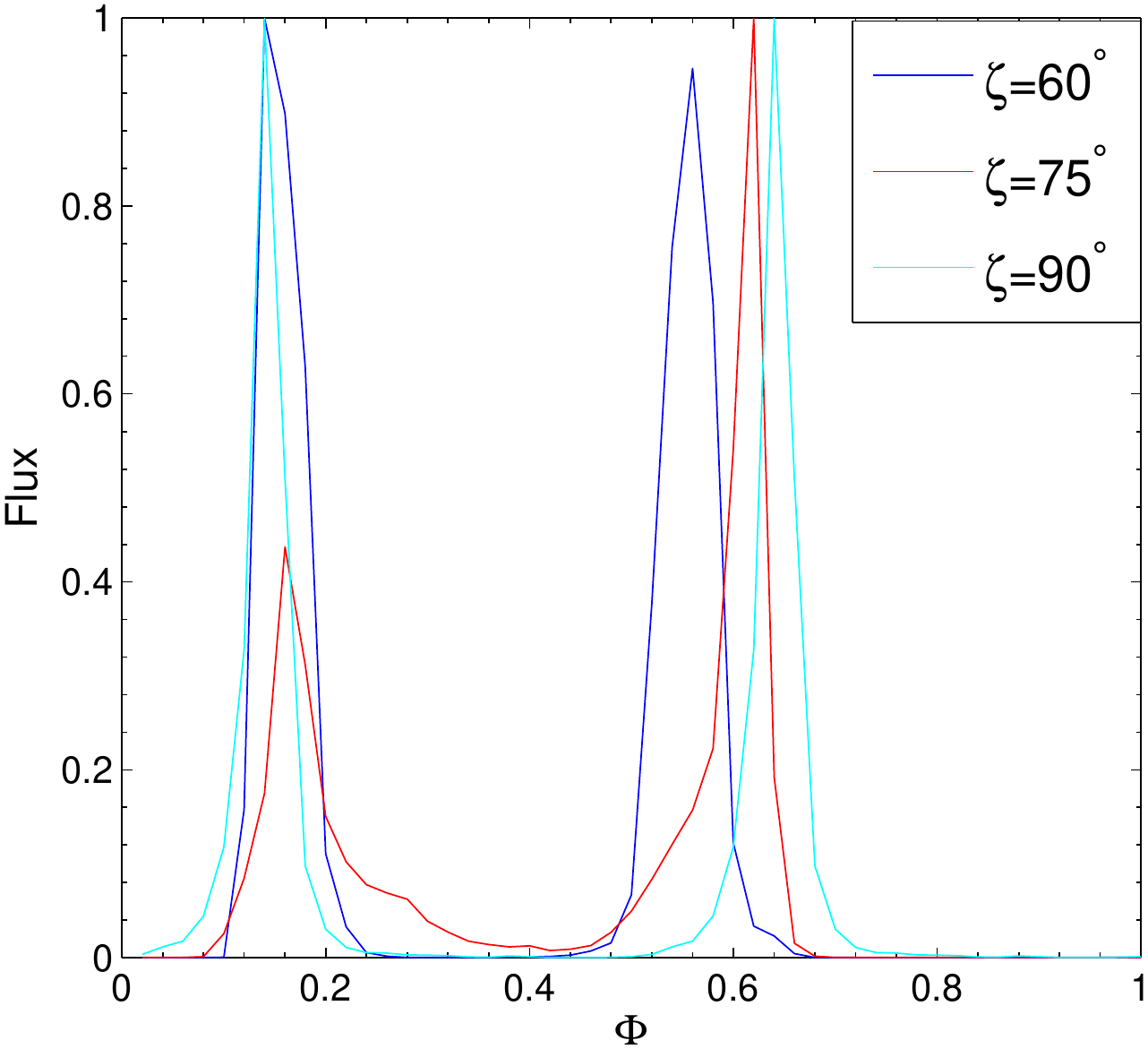} \qquad
\includegraphics[width=5.25 cm,height=5.35 cm]{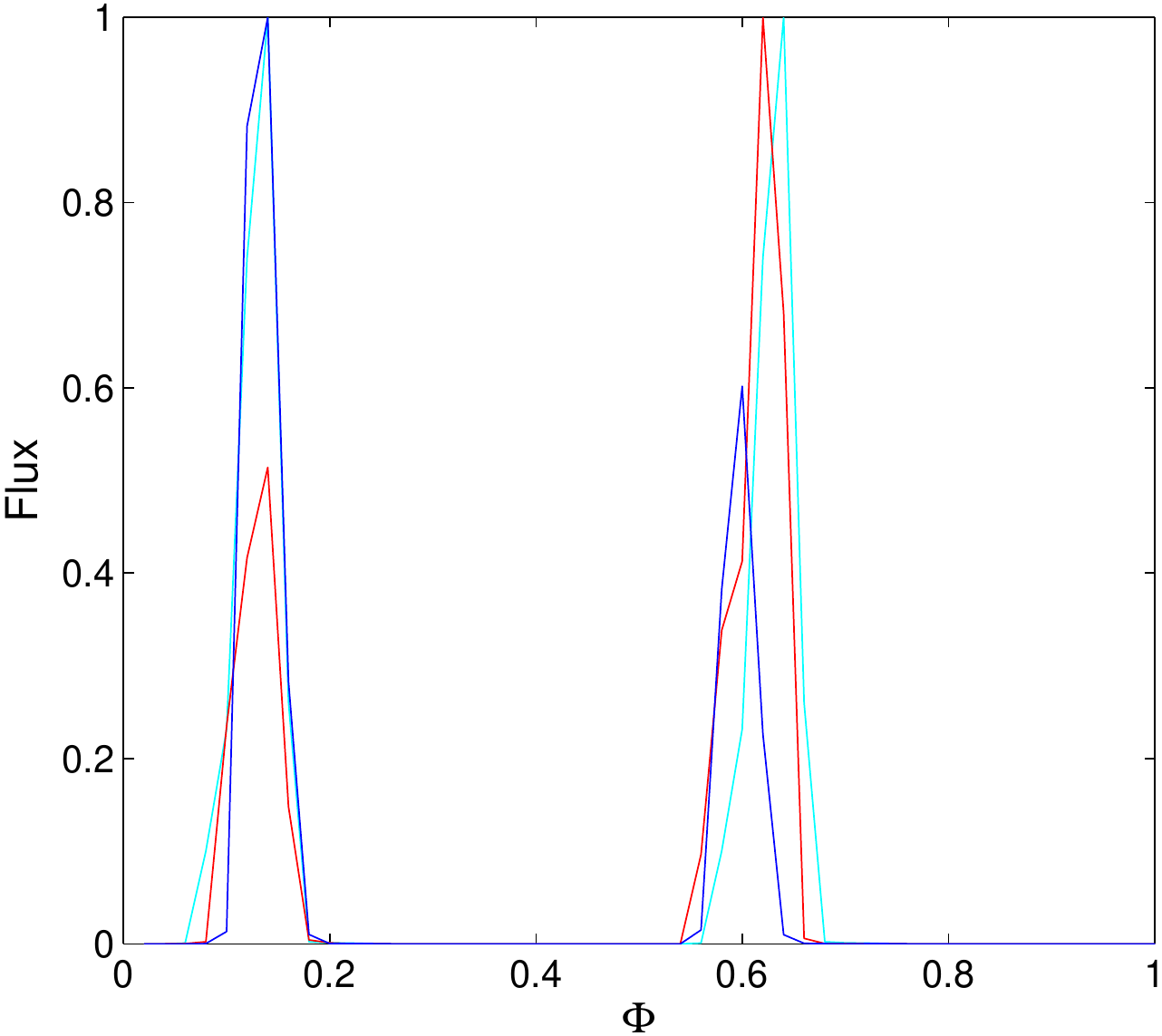} \qquad
\includegraphics[width=5.25 cm,height=5.35 cm]{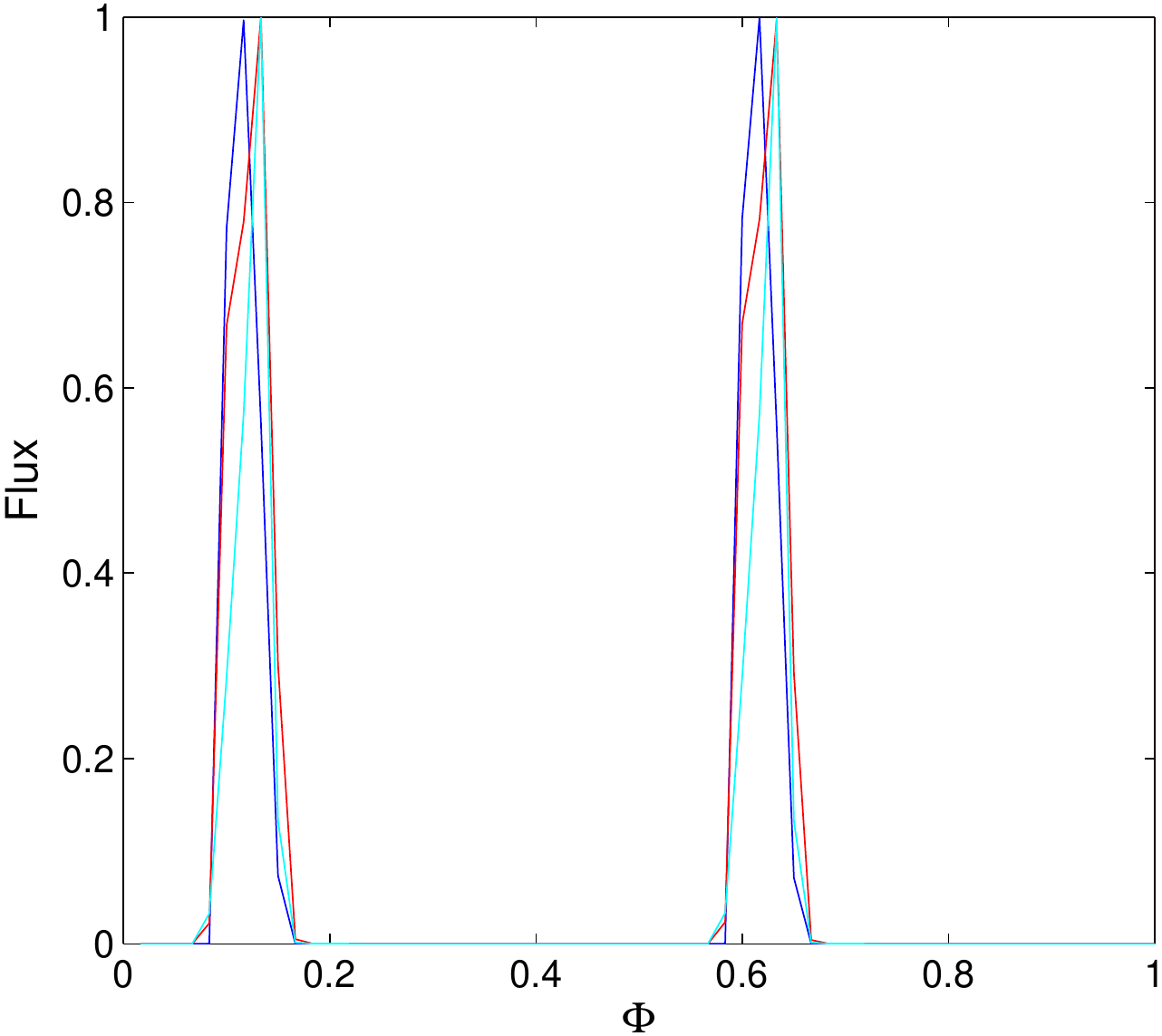} \qquad
\end{tabular}
\caption{The sky maps and the corresponding light curves at $>$ $1 \, \rm GeV$ energies in different inclination angles and view angles  with the pair multiplicity $\kappa=3$.  }
\label{Fig6}
\end{figure*}

A spectral filter is necessary to prevent  Gibbs oscillation and non-linear instabilities in the spectral algorithm. A non-physical dissipation is introduced by the filtering processes.
It is difficult to correctly catch the discontinuity induced by the  current sheet, which can be circumvent by increasing  simulation resolution and adjusting the spectral filter.
We show  the distribution of the magnetic field line and the accelerating electric field $E_0$ in the $x$$-$$z$  plane for  a $60^\circ$ rotator  with the pair multiplicity $\kappa=1$ by using different filter parameters in Figure \ref{Fig1}.
We see that all solutions have similar $E_0$ distributions in the current sheet, they are independent of the choice of a filter as it should be.
A low-order filter with ($\alpha,\beta$)=(10,6) gives more dissipation solution with large numerical diffusion  in the current sheet than a high-order filter with ($\alpha,\beta$)=(10,8) and  ($\alpha,\beta$)=(36,8). A more accurate solution can be obtained by using a optimized filtering parameter ($\alpha,\beta$)=(10,8) with a high-order filter and a suitable $\alpha$ value.
The normalized Poynting flux $L/L_{\rm aligned}$ as a function of radius $r$ for a  $60^\circ$ rotator with different pair multiplicities  are shown in Figure \ref{Fig2}.
For comparison, we also show the normalized Poynting flux with the pair multiplicity $\kappa=1$ for the low-resolution simulation  and the same optimized filtering parameter as the red dashed line.
It is found that  the Poynting flux increases with increasing $\kappa$ values  and approaches the force-free solution for the high $\kappa$ value. It is expected that there is no dissipation inside the LC in the combined FFE and AE magnetosphere. We see that the high-resolution simulation show smaller  dissipation in the LC compared to the low-resolution simulation. Our simulation show that the dissipative  region in the current sheet can be better resolved by both increasing the grid resolution and controlling the filtering effect in our spectral algorithm. It is noted that the low-resolution simulation and/or the high-resolution simulation with the strong spectral filter show more dissipative accelerating regions in the current sheet, which produces the higher cut-off energy in the curvature radiation spectrum.

We show  the distribution of the magnetic field line and the accelerating electric field $E_0$  for  a $60^\circ$ rotator with the pair multiplicity $\kappa=\{0,1,3\}$ and the optimized filtering parameters ($\alpha,\beta$)=(10,8) in the $x$$-$$z$ and $x$$-$$y$ plane in Figure \ref{Fig3}. As the pair multiplicity $\kappa$ increases,  the magnetosphere tends to  the force-free solution with a current sheet outside the LC. We observe a strong $E_0$ region with $E>B$  outside the LC.  The dissipative region decreases with increasing pair multiplicity and the dissipative region are more confined to near the current sheet outside the LC for high pair multiplicity. We also show the distribution of the magnetic field line and the accelerating electric field $E_0$  for  different inclination angles with the pair multiplicity $\kappa=3$  in the $x$$-$$z$  plane in Figure \ref{Fig4}. We see that all solutions have a near force-free magnetosphere with the dissipative region only near the current sheet for all the inclination angles. Our high-resolution simulation with  the optimized filtering parameter  gives more accurate solution near the current sheet compared to those in \citet{cao20} and \citet{pet20b}.\\

\subsection{Light curves and energy spectra}
The particle velocity in the radiation reaction limit is defined by \citep{gru12,gru13}
\begin{eqnarray}
{\bf v_{\pm}}=  {{\bf E} \times {\bf B}\pm(B_0{\bf {B}}+E_0{\bf {E}}) \over B^2+E^2_{0}},
\label{Eq6}
\end{eqnarray}
where the two signs correspond to positrons and  electrons, they  have a differen way to radiation reaction in the AE magnetosphere.
The Lorentz factor along particle  trajectory is given by
\begin{eqnarray}
\frac{d\gamma}{dt}=\frac{q_{\rm e}c E_{0}}{m_{\rm e}c^2}- \frac{2q^2_{\rm e} \gamma^4}{3R^2_{\rm CR}m_{\rm e}c} ,
\end{eqnarray}
The  photon spectrum of curvature radiation for each particle with Lorentz factor $\gamma$ is given
\begin{eqnarray}
F(E_{\gamma},r)=\frac{\sqrt{3} e^2 \gamma}{2 \pi \hbar R_{\rm CR} E_{\gamma}}F(x)\;,
\end{eqnarray}
where  $x=E_{\gamma}/E_{\rm cur}$, $E_{\gamma}$ is the  radiation photon energy, $E_{\rm cur}=\frac{3}{2}c\hbar \frac{\gamma^3}{R_{\rm CR}}$ is the characteristic energy of
the curvature radiation photon, $R_{\rm CR}$ is the curvature radius of particles, and the function $F(x)$ is defined as
\begin{equation}
F(x)=x\int_{x}^{\infty}{K_{\rm 5/3}}(\xi)\;d\xi,
\end{equation}

The initial particles are randomly  injected from the polar caps on the stellar surface with small Lorentz factor. The particle trajectory is determined by integrating the particle velocity from the the neutron star surface up to $r=2.5 \, r_{\rm L}$. The Lorentz factor along each trajectory is then computed under the influence of the local accelerating electric field and the curvature radiation loss. Assuming that the direction of the photon emission is along the direction of particle motion,  we can compute  the direction of the photon emission and the curvature radiation spectrum along each trajectory. The pulsar light curves and spectra can be constructed by collecting all curvature radiation photons from each radiating particle in sky maps.

We show  the curvature radiation spectra  for  a $60^\circ$ magnetosphere with different pair multiplicity $\kappa=\{0,1,3\}$ in Figure \ref{Fig5}.
It can be seen  that our model can produce a power-law spectrum with an exponential cut-off. Moreover, the cut-off energy decreases as the  pair multiplicity $\kappa$ increases, which is caused by the decrease of the accelerating electric field in the dissipative region outside the LC. We also find that the cut-off energy lies in the range of 1-5 GeV  for the pair multiplicity $\kappa\gtrsim1$, which is consistent with  the Fermi observed GeV cut-off energy. We also find that the spectra for all the inclination angles are  very similar to those presented in Figure \ref{Fig5}. We show the sky maps and the corresponding light curves in different inclination angles and view angles with the pair multiplicity $\kappa=3$ in Figure \ref{Fig6}. We see that the double-peak light curves can generally be produced for a broad range of the inclination angles and view angles, which are generally agreement with those observed by Fermi-LAT. Our results support that the observed $\gamma$-ray pulsar population is consistent with curvature radiation \citep{kal19}.

\section{Discussion and Conclusions}
We firstly explore the  properties of the pulsar light curves and energy spectra in dissipative AE magnetospheres. The dissipative AE magnetospheres with non-zero pair multiplicity are presented by a pseudo-spectral method with the high-resolution simulations in the co-moving frame. Our simulations show that the dissipative region near the current sheet outside the LC can  be accurately  captured by the high-resolution simulation. We use these field structures to define the trajectory of the positron and electron with the radiation reaction. The pulsar light curves and energy spectra are then produced by collecting all curvature radiation photons  along each particle trajectory. Our results show that the double-peak light curves and the power-law energy spectra with an exponential cut-off at $\sim \rm 1 \, GeV$ energy range can generally be produced for a moderate pair multiplicity $\kappa\gtrsim1$, which are well consistent with those observed by Fermi-LAT.

Our study provide the first step to model the pulsar emission for direct comparison with observation by including  consistent accelerating electric field and radiation reaction from the dissipative AE magnetosphere. It is necessary to perform more magnetosphere simulations with a broader range of the magnetic inclination and the pair multiplicity to construct the pulsar light curves and  spectra, which allow us directly compare them with the Fermi observations. This comparison will provide the meaningful constraints on the model parameters and enhance our understanding about the physical mechanisms of pulsar gamma-ray emission. We will present the detailed comparisons with the Fermi observational data in the near future.

\acknowledgments
We thank the anonymous referee for valuable comments and suggestions. We would like to thank  J\'{e}r$\hat{\rm o}$me P\'{e}tri and Ioannis Contopoulos for some useful discussions.  We acknowledge the financial support from the National Natural Science Foundation of China 12003026, and the Basic research Program of Yunnan Province 202001AU070070.


\end{document}